\documentclass[journal]{IEEEtran}
\usepackage{graphicx}
\usepackage[cmex10]{amsmath}
\usepackage{amsfonts,amssymb,latexsym,cite,ifthen,accents,multirow,setspace,wasysym}
\usepackage[usenames,dvipsnames]{color}
\usepackage[normalem]{ulem}
\usepackage[dvips,bookmarks]{hyperref}
\usepackage[all]{hypcap}
\usepackage[font=footnotesize]{caption}
\hypersetup{linkbordercolor=0 0 1, pdfauthor={J. Ziniel and P. Schniter}, pdftitle={Linear Binary Classification and Feature Selection via Generalized Approximate Message Passing}, pdfsubject={Linear binary classification with feature selection}, pdfkeywords={binary classification, feature selection, structured sparsity, structured feature selection, approximate message passing, compressed sensing}}
\usepackage{algorithm,algorithmic}
\usepackage{subfig}
\usepackage{breakurl}
\usepackage[11pt]{moresize}
\definecolor{darkgreen}{RGB}{0,77,15}
\definecolor{darkred}{RGB}{117,2,2}
\definecolor{lightgray}{RGB}{171,167,167}
\newboolean{ONE_COLUMN}\setboolean{ONE_COLUMN}{false}

 \newcommand{\defn}{\triangleq}

 \newcommand{\tvec}[1]{\ensuremath{\Tilde{\boldsymbol{#1}}}}
 
 \newcommand{\hvec}[1]{\ensuremath{\Hat{\boldsymbol{#1}}}}
 
 \renewcommand{\vec}[1]{\ensuremath{\boldsymbol{#1}}}
 \newcommand{\mat}[1]{\ensuremath{\begin{bmatrix}#1\end{bmatrix}}}

 \newcommand{\mc}[1]{\ensuremath{\mathcal{#1}}}
 \newcommand{\st}{{~\text{s.t.}~}}

 \newcommand{\Real}{{\mathbb{R}}}

 \newcommand{\giv}{\,|\,}
 \newcommand{\biggiv}{\,\big|\,}
 
 \newcommand{\tran}{^{\textsf{T}}}

 \DeclareMathOperator{\sgn}{sgn}
 
 \DeclareMathOperator{\var}{var}

 \DeclareMathOperator*{\argmax}{arg\, max}
 \DeclareMathOperator*{\argmin}{arg\, min}

 \DeclareMathOperator{\prox}{prox}

 \renewcommand{\eqref}[1]{(\ref{eq:#1})}
 \newcommand{\coderef}[1]{\ref{code:#1}}
 
 \newcommand{\Figref}[1]{Figure~\ref{fig:#1}}
 \newcommand{\figref}[1]{Fig.~\ref{fig:#1}}
 \newcommand{\tabref}[1]{Table~\ref{tab:#1}}
 \newcommand{\algref}[1]{Algorithm~\ref{alg:#1}}
 \newcommand{\secref}[1]{Section~\ref{sec:#1}}
 \newcommand{\Secref}[1]{Section~\ref{sec:#1}}
 \newcommand{\appref}[1]{Appendix~\ref{app:#1}}

 \newcommand{\textb}[1]{\textcolor{black}{#1}}
\renewcommand{\color}[1]{}

 \newcounter{comment}[section]
 
 \newcounter{texthead}[section]

  \floatstyle{plain}
  \newfloat{myalgo}{tbhp}{mya}
  \newenvironment{Algorithm}[2][tbh]%
  {\begin{myalgo}[#1]
  \centering
  \begin{minipage}{#2}
  \begin{algorithm}[H]}%
  {\end{algorithm}
  \end{minipage}
  \end{myalgo}}

  \newcommand{\ubar}[1]{\underaccent{\bar}{#1}}
  \newcommand{\eps}{\varepsilon}

  \renewcommand{\var}[1]{\text{var}\{#1\}}

  \newcommand{\E}{\textsf{e}} 
  \newcommand{\Y}{\textsf{y}} 
  \newcommand{\Z}{\textsf{z}} 
  \newcommand{\W}{\textsf{w}} 
  \newcommand{\vE}{\textsf{\textbf{\textit{e}}}} 
  \newcommand{\vY}{\textsf{\textbf{\textit{y}}}} 
  \newcommand{\vZ}{\textsf{\textbf{\textit{z}}}} 
  \newcommand{\vW}{\textsf{\textbf{\textit{w}}}} 
  \newcommand{\vX}{\textsf{\textbf{\textit{x}}}} 
  \newcommand{\vXX}{\textsf{\textbf{\textit{X}}}} 
 
 \newcounter{threshcounter}

 \newcounter{algcounter}

 \newcounter{emcounter}

 \graphicspath{{./}{Figures/}}

\begin{document}
\setlength{\arraycolsep}{0.5mm}

%
\title{Binary Linear Classification and Feature Selection via Generalized Approximate Message Passing}
\author{Justin Ziniel, Philip Schniter,\IEEEauthorrefmark{1} and Per Sederberg
\thanks{Ziniel and Schniter are with the Dept.\ of Electrical and Computer Engineering, The Ohio State University, Columbus, Ohio; e-mail: \{zinielj, schniter\}@ece.osu.edu.  Their work on this project has been supported by NSF grant CCF-1218754, by NSF grant CCF-1018368, by DARPA/ONR grant N66001-10-1-4090, and by an allocation of computing time from the Ohio Supercomputer Center.}
\thanks{Sederberg is with the Dept. of Psychology, The Ohio State University, Columbus, Ohio; e-mail: sederberg.1@osu.edu.}
\thanks{\IEEEauthorrefmark{1}Please direct all correspondence to Prof. Philip Schniter, Dept. ECE, The Ohio State University, 2015 Neil Ave., Columbus OH 43210, e-mail: schniter@ece.osu.edu, phone 614.247.6488, fax 614.292.7596.}
\thanks{Portions of this work were presented at the 2013 Workshop on Information Theory and its Applications in San Diego, CA, and the 2014 Conference on Information Sciences and Systems in Princeton, NJ.}
}

\maketitle

\begin{abstract}
For the problem of binary linear classification and feature selection, we propose algorithmic approaches to classifier design based on the generalized approximate message passing (GAMP) algorithm, recently proposed in the context of compressive sensing.
We are particularly motivated by problems where the number of features greatly exceeds the number of training examples, but where only a few features suffice for accurate classification.
We show that sum-product GAMP can be used to (approximately) minimize the classification error rate and max-sum GAMP can be used to minimize a wide variety of regularized loss functions.
Furthermore, we describe an expectation-maximization (EM)-based scheme to learn the associated model parameters online, as an alternative to cross-validation, and we show that GAMP's state-evolution framework can be used to accurately predict the misclassification rate.
Finally, we present a detailed numerical study to confirm the accuracy, speed, and flexibility afforded by our GAMP-based approaches to binary linear classification and feature selection.
\end{abstract}

\section{Introduction}
\label{sec:introduction}

In this work we consider binary linear classification and feature selection \cite{B2006}.  
The objective of \emph{binary linear classification} is to learn the weight vector $\vW \in \mathbb{R}^N$ that best predicts an unknown binary class label $\Y \in \{-1,1\}$ associated with a given vector of quantifiable features $\vec{x} \in \mathbb{R}^N$ from the sign of a linear ``score'' $\Z \triangleq \langle \vec{x}, \vW \rangle$.\footnote{We note that one could also compute the score from a fixed non-linear transformation $\psi(\cdot)$ of the original feature $\vec{x}$ via $\Z \triangleq \langle \psi(\vec{x}), \vW \rangle$ as in kernel-based classification.  Although the methods we describe here are directly compatible with this approach, we write $\Z=\langle \vec{x}, \vW \rangle$ for simplicity.}  
The goal of \emph{linear feature selection} is to identify which subset of the $N$ weights in $\vW$ are necessary for accurate prediction of the unknown class label $\Y$, since in some applications (e.g., multi-voxel pattern analysis) this subset itself is of primary concern.  

In formulating this linear feature selection problem, we assume that there exists a $K$-\emph{sparse} weight vector $\vW$ (i.e., $\|\vW\|_0 = K \ll N$) such that
$\Y = \sgn(\langle \vec{x}, \vW \rangle - \E)$,
where $\sgn(\cdot)$ is the signum function and $\E \sim p_{\E}$ is a random perturbation accounting for model inaccuracies.
For the purpose of learning $\vW$, we assume the availability of $M$ labeled training examples generated independently according to this model:
\begin{equation}
        y_m = \sgn(\langle \vec{x}_m, \vW \rangle - \E_m),      \quad \forall \, m = 1,\ldots,M,
        \label{eq:training_model}
\end{equation}
with $\E_m \sim \text{i.i.d } p_{\E}$.  
It is common to express the relationship between the label $y_m$ and the score $\Z_m \triangleq \langle \vec{x}_m, \vW \rangle$ in \eqref{training_model} via the conditional pdf $p_{\Y_m|\Z_m}(y_m | z_m)$, known as the ``activation function,'' which can be related to the perturbation pdf $p_{\E}$ via
\begin{equation}
        p_{\Y_m|\Z_m}(1 | z_m) 
        = \int_{-\infty}^{z_m} p_{\E}(e) \, de 
        = 1 - p_{\Y_m|\Z_m}(-1 | z_m) 
        \label{eq:activation_integral} .
\end{equation}

We are particularly interested in classification problems in which the number of potentially discriminatory features $N$ drastically exceeds the number of available training examples $M$.  Such computationally challenging problems are of great interest in a number of modern applications, including text classification \cite{F2003b}, multi-voxel pattern analysis (MVPA) \cite{HGFISP2001, NPDH2006,PMB2009}, 
conjoint analysis \cite{GHH2007}, and micro-array gene expression \cite{XJK2001}.  In MVPA, for instance, neuro-scientists attempt to infer which regions in the human brain are responsible for distinguishing between two cognitive states by measuring neural activity via fMRI at $N \sim 10^4$ voxels.  Due to the expensive and time-consuming nature of working with human subjects, classifiers are routinely trained using only $M \sim 10^2$ training examples, and thus $N\gg M$.

In the $N \gg M$ regime, the model of \eqref{training_model} coincides with that of \emph{noisy one-bit compressed sensing} (CS) \cite{BB2008,PV2013}.  
In that setting, it is typical to write \eqref{training_model} in matrix-vector form using $\vec{y} \triangleq [y_1,\ldots,y_M]\tran$, $\vE \triangleq [\E_1,\ldots,\E_M]\tran$, $\vec{X} \triangleq [\vec{x}_1,\ldots,\vec{x}_M]\tran$, and element-wise $\sgn(\cdot)$, yielding
\begin{equation}
        \vec{y} = \sgn(\vec{X}\vW - \vE),
        \label{eq:matrix_training_model}
\end{equation}
where $\vW$ embodies the signal-of-interest's sparse representation, $\vec{X} = \vec{\Phi} \vec{\Psi}$ is a concatenation of a linear measurement operator $\vec{\Phi}$ and a sparsifying signal dictionary $\vec{\Psi}$, and $\vE$ is additive noise.\footnote{For example, the common case of additive white Gaussian noise (AWGN) $\{\E_m\} \sim \text{i.i.d } \mathcal{N}(0,v)$ corresponds to the ``probit'' activation function, i.e., $p_{\Y_m|\Z_m}(1 | z_m) = \Phi(z_m/v)$, where $\Phi(\cdot)$ is the standard-normal cdf.} 
Importantly, in the $N \!\gg\! M$ setting, \cite{PV2013} established performance guarantees on the estimation of $K$-sparse $\vW$ from $O(K \log N/K)$ binary measurements of the form \eqref{matrix_training_model}, under i.i.d Gaussian $\{\vec{x}_m\}$ and mild conditions on the perturbation process $\{\E_m\}$, even when the entries within $\vec{x}_m$ are correlated.  
This result implies that, in large binary linear classification problems, accurate feature selection is indeed possible from $M \ll N$ training examples, as long as the underlying weight vector $\vW$ is sufficiently sparse. 
Not surprisingly, many techniques have been proposed to find such weight vectors \cite{KS1996,KJ1997,T2001,F2001,F2003,K2007,CTY2009,YCHL2010}.

In addition to theoretical analyses, the CS literature also offers a number of high-performance algorithms for the inference of $\vW$ in \eqref{matrix_training_model}, e.g., \cite{BB2008,GNR2010,LWYB2011,KBAU2012,KGR2012,PV2013}.  
Thus, the question arises as to whether these algorithms also show advantages in the domain of binary linear classification and feature selection.  
In this paper, we answer this question in the affirmative by focusing on the \emph{generalized approximate message passing} (GAMP) algorithm \cite{R2011}, which extends the AMP algorithm \cite{DMM2009,DMM2010} from the case of linear, AWGN-corrupted observations (i.e., $\vec{y}=\vec{X}\vW-\vE$ for $\vE\sim\mc{N}(\vec{0},v\vec{I})$) to the case of generalized-linear observations, such as \eqref{matrix_training_model}.  
AMP and GAMP are attractive for several reasons: 
\textit{(i)} For i.i.d sub-Gaussian $\vec{X}$ in the large-system limit (i.e., $M,N\rightarrow\infty$ with fixed ratio $\delta=\frac{M}{N}$), they are rigorously characterized by a state-evolution whose fixed points, when unique, are optimal \cite{Javanmard:II:13}; 
\textit{(ii)} Their state-evolutions predict fast convergence rates; 
\textit{(iii)} They are very flexible with regard to data-modeling assumptions (see, e.g., \cite{ZRS2012}); 
\textit{(iv)} Their model parameters can be learned online using an expectation-maximization (EM) approach that has been shown to yield state-of-the-art mean-squared reconstruction error in CS problems \cite{VS2013}.

In this work, we develop a GAMP-based approach to binary linear classification and feature selection that makes the following contributions:
1) in \secref{gamp}, we show that GAMP implements a particular approximation to the error-rate minimizing linear classifier under the assumed model \eqref{training_model};
2) in \secref{state_evolution}, we show that GAMP's state evolution framework can be used to characterize the misclassification rate in the large-system limit;
3) in \secref{gamp_classification}, we develop methods to implement logistic, probit, and hinge-loss-based regression using both max-sum and sum-product versions of GAMP, and we further develop a method to make these classifiers robust in the face of corrupted training labels; and
4) in \secref{parameter_tuning}, we present an EM-based scheme to learn the model parameters online, as an alternative to cross-validation. 
The numerical study presented in \secref{numerical_study} then confirms the efficacy, flexibility, and speed afforded by our GAMP-based approaches to binary classification and feature selection.

\emph{Notation}:
Random quantities are typeset in sans-serif (e.g., $\E$) while deterministic quantities are typeset in serif (e.g., $e$).  
The pdf of random variable $\E$ under deterministic parameters $\vec{\theta}$ is written as $p_{\E}(e;\vec{\theta})$, where the subscript and parameterization are sometimes omitted for brevity.
Column vectors are typeset in boldface lower-case (e.g., $\vec{y}$ or $\vY$), matrices in boldface upper-case (e.g., $\vec{X}$ or $\vXX$), and their transpose is denoted by $(\cdot)\tran$.
For vector $\vec{y}=[y_1,\dots,y_N]\tran$, $\vec{y}_{m:n}$ refers to the subvector $[y_m,\dots,y_n]\tran$.
Finally, $\mathcal{N}(\vec{a};\vec{b},\vec{C})$ is the multivariate normal distribution as a function of $\vec{a}$, with mean $\vec{b}$, and with covariance matrix $\vec{C}$,
while $\phi(\cdot)$ and $\Phi(\cdot)$ denote the standard normal pdf and cdf, respectively.

\section{GAMP for Classification}
\label{sec:gamp}
In this section, we introduce generalized approximate message passing (GAMP) from the perspective of binary linear classification.  In particular, we show that the \emph{sum-product} variant of GAMP is a loopy belief propagation (LBP) approximation of the classification-error-rate minimizing linear classifier and that the \emph{max-sum} variant of GAMP is a LBP implementation of the standard regularized-loss-minimization approach to linear classifier design.

\subsection{Sum-Product GAMP}
\label{sec:gamp:sum_prod}
Suppose that we are given $M$ labeled training examples $\{y_m,\vec{x}_m\}_{m=1}^M$, and $T$ test feature vectors $\{\vec{x}_t\}_{t=M+1}^{M+T}$ associated with unknown test labels $\{\Y_t\}_{t=M+1}^{M+T}$, all obeying the noisy linear model \eqref{training_model} under some known error pdf $p_{\E}$, and thus known $p_{\Y_m|\Z_m}$.  
We then consider the problem of computing the classification-error-rate minimizing hypotheses $\{\hat{y}_t\}_{t=M+1}^{M+T}$,
\begin{align}
\hat{y}_t 
&= \argmax_{y_t\in\{-1,1\}} p_{\Y_t|\vY_{1:M}}\big(y_t\biggiv\vec{y}_{1:M};\vec{X}\big),
\label{eq:optimal_label}
\end{align}
with $\vec{y}_{1:M}\defn[y_1,\dots,y_M]\tran$ and $\vec{X}\defn[\vec{x}_1,\dots,\vec{x}_{M+T}]\tran$.  Note that we treat the labels $\{\Y_m\}_{m=1}^{M+T}$ as random but the features $\{\vec{x}_m\}_{m=1}^{M+T}$ as deterministic parameters.  The probabilities in \eqref{optimal_label} can be computed via the marginalization
\begin{align}
\lefteqn{ p_{\Y_t|\vY_{1:M}}\big(y_t\biggiv\vec{y}_{1:M};\vec{X}\big)
= p_{\Y_t,\vY_{1:M}}\big(y_t,\vec{y}_{1:M};\vec{X}\big) C_{\vY}^{-1} }\\
&= C_{\vY}^{-1} \sum_{\vec{y}\in\mathcal{Y}_t(y_t)} \int p_{\vY,\vW}(\vec{y},\vec{w};\vec{X}) \,d\vec{w} \hspace{1in}
        \label{eq:marginalize_labels}
\end{align}
with scaling constant $C_{\vY}\defn p_{\vY_{1:M}}\big(\vec{y}_{1:M};\vec{X}\big)$, label vector $\vec{y}=[y_1,\dots,y_{M+T}]\tran$, and constraint set $\mathcal{Y}_t(y)\defn\{\tvec{y}\in\{-1,1\}^{M+T}\st [\tvec{y}]_t=y \text{~and~}[\tvec{y}]_m=y_m~\forall m=1,\dots,M\}$ which fixes the $t$th element of $\vec{y}$ at the value $y$ and the first $M$ elements of $\vec{y}$ at the values of the corresponding training labels.  The joint pdf in \eqref{marginalize_labels} factors as
\begin{align}
p_{\vY,\vW}(\vec{y},\vec{w};\vec{X})
&= \prod_{m=1}^{M+T} p_{\Y_m|\Z_m}(y_m\giv \vec{x}_m\tran\vec{w})\, \prod_{n=1}^{N}p_{\W_n}\!(w_n) 
        \label{eq:predict_factor_rep}
\end{align}
due to the model \eqref{training_model} and assuming a separable prior, i.e.,
\begin{equation}
        p_{\vW}(\vec{w}) = \prod_{n=1}^N p_{\W_n}\!(w_n).
        \label{eq:separable_prior}
\end{equation}
Although the separability assumption can also be relaxed (see, e.g., \cite{S2010a,ZRS2012}), we do not consider such extensions in this work.

\begin{figure}
\centering
\subfloat[][Full]{\scalebox{0.875}{\includegraphics*[0.85in,9.45in][2.72in,11.75in]{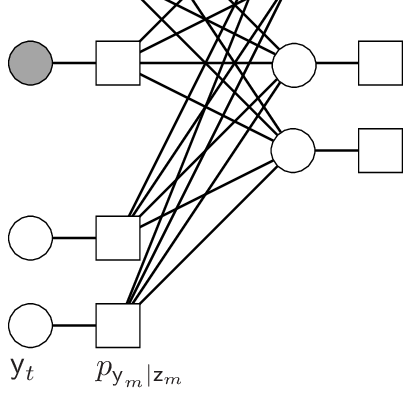}}\label{fig:full_factor_graph}}
\quad
\subfloat[][Reduced]{\scalebox{0.875}{\includegraphics*[0.85in,9.85in][2.72in,11.75in]{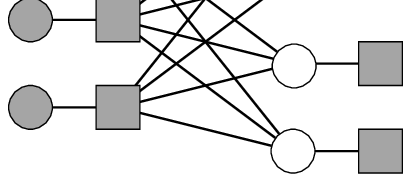}}\label{fig:reduced_factor_graph}}
\caption{Factor graph representations of the integrand of \eqref{predict_factor_rep}, with white/grey circles denoting unobserved/observed random variables, and rectangles denoting pdf ``factors''.}
\label{fig:discriminative_factor_graph}
\end{figure}

The factorization \eqref{predict_factor_rep} is illustrated using the \emph{factor graph} in \figref{full_factor_graph}, which connects the various random variables to the pdf factors in which they appear.  Although exact computation of the marginal posterior test-label probabilities via \eqref{marginalize_labels} is computationally intractable due to the high-dimensional summation and integration, the factor graph in \figref{full_factor_graph} suggests the use of loopy belief propagation (LBP) \cite{FM1998}, and in particular the \emph{sum-product algorithm} (SPA) \cite{KFL2001}, as a tractable way to approximate these marginal probabilities.  Although the SPA guarantees exact marginal posteriors only under non-loopy (i.e., tree-structured graphs), it has proven successful in many applications with loopy graphs, such as turbo decoding \cite{MMC1998}, computer vision \cite{FPC2000}, and compressive sensing \cite{DMM2009,DMM2010,R2011}.

Because a direct application of the SPA to the factor graph in \figref{full_factor_graph} is itself computationally infeasible in the high-dimensional case of interest, we turn to a recently developed approximation: the sum-product variant of GAMP \cite{R2011}, as specified in \algref{gamp}.
The GAMP algorithm is specified in \algref{gamp} for a given instantiation of $\vec{X}$, $p_{\Y|\Z}$, and $\{p_{\W_n}\!\}$. 
There, the expectation and variance in lines \coderef{sum_prod_z}-\coderef{sum_prod_tz} and \coderef{sum_prod_w}-\coderef{sum_prod_tw} are taken element-wise w.r.t the GAMP-approximated marginal posterior pdfs (with superscript $k$ denoting the iteration)
\begin{align}
q(z_m\giv \hat{p}_m^k,\tau_{p_m}^k)
&= p_{\Y_m|\Z_m}(y_m|z_m)\,\mathcal{N}(z_m;\hat{p}_m^k,\tau_{p_m}^k) \,C_{\Z}^{-1}      
\label{eq:z_posterior} \\
q(w_n\giv \hat{r}_n^k,\tau_{r_n}^k)
&= p_{\W_n}\!(w_n)\,\mathcal{N}(w_n;\hat{r}_n^k,\tau_{r_n}^k) \,C_{\W}^{-1}
\label{eq:w_posterior} 
\end{align}
with appropriate normalizations $C_{\Z}$ and $C_{\W}$, and the vector-vector multiplications and divisions in lines \coderef{p}, \coderef{max_sum_tz}, \coderef{ts}, \coderef{s}, \coderef{tr}, \coderef{r}, \coderef{max_sum_tw} are performed element-wise.
Due to space limitations, we refer the interested reader to \cite{R2011} for an overview and derivation of GAMP, to \cite{Javanmard:II:13} for rigorous analysis under large i.i.d sub-Gaussian $\vec{X}$, and to \cite{RSRFC2013,RSF2014} for fixed-point and local-convergence analysis under arbitrary $\vec{X}$.

\ifthenelse{\boolean{ONE_COLUMN}}
{\begin{Algorithm}{3.5in}}
{\begin{Algorithm}{3.5in}}
        \caption{Generalized Approximate Message Passing}
        \label{alg:gamp}
        \singlespacing
        \begin{algorithmic}[1]
                \REQUIRE{Matrix $\vec{X}$, priors $p_{\W_n}\!(\cdot)$, activation functions $p_{\Y_m|\Z_m}(y_m|\cdot)$, and mode $\in\{\texttt{SumProduct},\texttt{MaxSum}\}$}
                \ENSURE{$k \leftarrow 0$;\, $\hvec{s}^{-1} \!\leftarrow\! \vec{0}$;\, $\vec{S} \!\leftarrow\! |\vec{X}|^2$;\, $\hvec{w}^0 \!\leftarrow\! \vec{0}$;\, $\vec{\tau}_w^0 \!\leftarrow\! \vec{1}$}
                \REPEAT
                        \STATE $\vec{\tau}_p^k \leftarrow \vec{S}\vec{\tau}_w^k$ \label{code:tp}
                        \STATE $\hvec{p}^k \leftarrow \vec{X}\hvec{w}^k - \hvec{s}^{k-1}\vec{\tau}_p^k$ \label{code:p}
                        \IF{\texttt{SumProduct}}
                                \STATE $\hvec{z}^k \leftarrow \text{E}\{\vZ \giv \hvec{p}^k, \vec{\tau}_p^k\}$  \label{code:sum_prod_z}
                                \STATE $\vec{\tau}_z^k \leftarrow \text{var}\{\vZ \giv \hvec{p}^k, \vec{\tau}_p^k\}$    \label{code:sum_prod_tz}
                        \ELSIF{\texttt{MaxSum}}
                                \STATE  $\hvec{z}^k \gets \prox_{\vec{\tau}_p^k f_{\Z_m}}\!(\hvec{p}^k)$ \label{code:max_sum_z}
                                \STATE $\vec{\tau}_z^k \gets \vec{\tau}_p^k \prox'_{\vec{\tau}_p^k f_{\Z_m}}\!(\hvec{p}^k)$ \label{code:max_sum_tz}

                        \ENDIF
                        \STATE $\vec{\tau}_s^k \leftarrow 1 / \vec{\tau}_p^k - \vec{\tau}_z^k / (\vec{\tau}_p^k)^2$ \label{code:ts}
                        \STATE $\hvec{s}^k \leftarrow (\hvec{z}^k - \hvec{p}^k) / \vec{\tau}_p^k$ \label{code:s}
                        \STATE $\vec{\tau}_r^k \leftarrow 1 / (\vec{S}^{\textsf{T}} \vec{\tau}_s^k)$ \label{code:r}
                        \STATE $\hvec{r}^k \leftarrow \hvec{w}^k + \vec{\tau}_r^k \vec{X}^{\textsf{T}}\hvec{s}^k$ \label{code:tr}
                        \IF{\texttt{SumProduct}}
                                \STATE $\hvec{w}^{k+1} \leftarrow \text{E}\{\vW \giv \hvec{r}^k, \vec{\tau}_r^k\}$      \label{code:sum_prod_w}
                                \STATE $\vec{\tau}_w^{k+1} \leftarrow \text{var}\{\vW \giv \hvec{r}^k, \vec{\tau}_r^k\}$        \label{code:sum_prod_tw}
                        \ELSIF{\texttt{MaxSum}}
                                \STATE  $\hvec{w}^{k+1} \gets \prox_{\vec{\tau}_r^k f_{\W_n}}\!(\hvec{r}^k)$ \label{code:max_sum_w}
                                \STATE $\vec{\tau}_w^{k+1} \gets \vec{\tau}_r^k \prox'_{\vec{\tau}_r^k f_{\W_n}}\!(\hvec{r}^k)$ \label{code:max_sum_tw}

                        \ENDIF
                        \STATE $k \leftarrow k + 1$
                \UNTIL{Terminated}
        \end{algorithmic}
\end{Algorithm}

Applying GAMP to the classification factor graph in \figref{full_factor_graph} and examining the resulting form of lines \coderef{sum_prod_z}-\coderef{sum_prod_tz} in \algref{gamp}, it becomes evident that the test-label nodes $\{\Y_t\}_{t=M+1}^{M+T}$ do not affect the GAMP weight estimates $(\hvec{w}^k,\vec{\tau}_w^k)$ and thus the factor graph can effectively be simplified to the form shown in \figref{reduced_factor_graph}, after which the (approximated) posterior test-label pdfs are computed via
\begin{equation}
p_{\Y_t|\vY_{1:M}}\!\big(y_t|\vec{y}_{1:M};\vec{X}\big)
\approx \int p_{\Y_t|\Z_t}\!(y_t|z_t)\,\mc{N}(z_t;\hat{z}^\infty_t,\tau_{z,t}^\infty)\,dz_t
\end{equation}
where $\hat{z}^\infty_t$ and $\tau_{z,t}^\infty$ denote the $t^{th}$ element of the GAMP vectors $\hvec{z}^k$ and $\vec{\tau}_z^k$, respectively, at the final iteration ``$k=\infty$.''

\subsection{Max-Sum GAMP}
\label{sec:gamp:max_sum}
An alternate approach to linear classifier design is through the minimization of a regularized loss function, e.g., 
\begin{equation}
        \hvec{w} = \argmin_{\vec{w}\in\Real^N} \sum_{m=1}^M f_{\Z_m}\!(\vec{x}_m\tran\vec{w}) + \sum_{n=1}^N f_{\W_n}\!(w_n),
        \label{eq:reg_loss_min}
\end{equation}
where $f_{\Z_m}\!(\cdot)$ are $y_m$-dependent convex loss functions (e.g., logistic, probit, or hinge based) and where $f_{\W_n}\!(\cdot)$ are convex regularization terms (e.g., $f_{\W_n}\!(w) = \lambda w^2$ for $\ell_2$ regularization and $f_{\W_n}\!(w) = \lambda |w|$ for $\ell_1$ regularization).

The solution to \eqref{reg_loss_min} can be recognized as the \emph{maximum a posteriori} (MAP) estimate of random vector $\vW$ 
given a separable prior $p_{\vW}(\cdot)$ and likelihood corresponding to \eqref{training_model}, i.e., 
\begin{equation}
p_{\vY|\vW}(\vec{y}|\vec{w};\vec{X})
=\prod_{m=1}^M p_{\Y_m|\Z_m}(y_m|\vec{x}_m\tran\vec{w})
\label{eq:separable_likelihood} ,
\end{equation}
when $f_{\Z_m}\!(z)=-\log p_{\Y_m|\Z_m}(y_m|z)$ and $f_{\W_n}\!(w)=-\log p_{\W_n}\!(w)$.
Importantly, this statistical model is exactly the one yielding the reduced factor graph in \figref{reduced_factor_graph}.

Similar to how sum-product LBP can be used to compute (approximate) marginal posteriors in loopy graphs, \emph{max-sum} LBP can be used to compute the MAP estimate \cite{L2004}.
Since max-sum LBP is itself intractable for the high-dimensional problems of interest, we turn to the max-sum variant of GAMP \cite{R2011}, which is also specified in \algref{gamp}.
There, lines \coderef{max_sum_z}-\coderef{max_sum_tz} are to be interpreted as 
\begin{align}
\hat{z}_m^k 
&= \prox_{\tau_{p_m}^k f_{\Z_m}}\!(\hat{p}_m^k),~~m=1,\dots,M,
\label{eq:prox}\\
\tau_{z_m}^k 
&= \tau_{p_m}^k\prox'_{\tau_{p_m}^k f_{\Z_m}}\!(\hat{p}_m^k),~~m=1,\dots,M,
\label{eq:proxderiv} 
\end{align}
with $(\cdot)'$ and $(\cdot)''$ denoting first and second derivatives and 
\begin{align} 
\prox_{\tau f}(v) 
&\defn \argmin_{u \in \Real} \Big[ f(u) + \frac{1}{2\tau}(u-v)^2 \Big] 
\label{eq:proxDef} \\
\prox'_{\tau f}(v) 
&= \big(1+\tau f''(\prox_{\tau f}(v))\big)^{-1} ,
\label{eq:proxderiv2} 
\end{align}
and lines \coderef{max_sum_w}--\coderef{max_sum_tw} are to be interpreted similarly.
It is known \cite{RSRFC2013} that, for \emph{arbitrary} $\vec{X}$, the fixed points of GAMP correspond to the critical points of the optimization objective \eqref{reg_loss_min}.

\subsection{GAMP Summary}

In summary, the sum-product and max-sum variants of the GAMP algorithm provide tractable methods of approximating the posterior test-label probabilities $\{p_{\Y_t|\vY_{1:M}}\!(y_t|\vec{y}_{1:M})\}_{t=T+1}^{M+T}$ and finding the MAP weight vector $\hvec{w}=\argmax_{\vec{w}} p_{\vW|\vY_{1:M}}\!(\vec{w}|\vec{y}_{1:M})$, respectively, under the label-generation model \eqref{separable_likelihood} [equivalently, \eqref{training_model}] and the separable weight-vector prior \eqref{separable_prior}, assuming that the distributions $p_{\Y|\Z}$ and $\{p_{\W_n}\!\}$ are known and facilitate tractable scalar-nonlinear update steps \coderef{sum_prod_z}-\coderef{sum_prod_tz}, \coderef{max_sum_z}-\coderef{max_sum_tz}, \coderef{sum_prod_w}-\coderef{sum_prod_tw}, and \coderef{max_sum_w}-\coderef{max_sum_tw}.
In \secref{gamp_classification}, we discuss the implementation of these update steps for several popular activation functions, and 
in \secref{parameter_tuning}, we discuss how the parameters of $p_{\Y_m|\Z_m}$ and $p_{\W_n}$ can be learned online. 

\section{Misclassification Rate via State Evolution}
\label{sec:state_evolution}
As mentioned earlier, the behavior of GAMP in the large-system limit (i.e., $M,N\rightarrow\infty$ with fixed ratio $\delta=\frac{M}{N}$) under i.i.d sub-Gaussian $\vec{X}$ is characterized by a scalar state evolution \cite{R2011,Javanmard:II:13}.  
We now describe how this state evolution can be used to characterize the test-error rate of the linear-classification GAMP algorithms described in \secref{gamp}.

The GAMP state evolution characterizes average GAMP performance over an ensemble of (infinitely sized) problems, each associated with one realization $(\vec{y},\vec{X},\vec{w})$ of the random triple $(\vY,\vXX,\vW)$. 
Recall that, for a given problem realization $(\vec{y},\vec{X},\vec{w})$, the GAMP iterations in \algref{gamp} yields the sequence of estimates $\{\hvec{w}^k\}_{k=1}^\infty$ of the true weight vector $\vec{w}$.
Then, according to the state evolution, 
$p_{\vW,\hat{\vW}^k}(\vec{w},\hvec{w}^k) \sim \prod_{n} p_{\W_n,\hat{\W}_n^k}(w_n,\hat{w}_n^k)$
and the first two moments of the joint pdf $p_{\W_n,\hat{\W}_n^k}$ can be computed using \cite[Algorithm 3]{R2011}.

Suppose that the $(\vY,\vXX)$ above represent training examples associated with a true weight vector $\vW$, and that $(\Y,\vX)$ represents a test pair also associated with the same $\vW$ and with $\vX$ having i.i.d elements distributed identically to those of $\vXX$ (with, say, variance $\frac{1}{M}$).
The true and iteration-$k$-estimated test scores are then $\Z \triangleq \vX^{\textsf{T}} \vW$ and $\hat{\Z}^k \triangleq \vX^{\textsf{T}} \widehat{\vW}^k$, respectively.
The corresponding test-error rate\footnote{For simplicity we assume a decision rule of the form $\hat{\Y}^k = \sgn(\hat{\Z}^k)$, although other decision rules can be accommodated in our analysis.}  
$\mathcal{E}^k \defn \text{Pr}\{\Y \neq \sgn(\hat{\Z}^k)\}$ can be computed as follows.
Letting $I_{\{\cdot\}}$ denote an indicator function that assumes the value $1$ when its Boolean argument is true and the value $0$ otherwise, we have
\begin{align}
\lefteqn{ \mathcal{E}^k = \text{E}\big\{I_{\{\Y \neq \sgn(\hat{\Z}^k)\}}\big\}} \\
                &= \!\!\!\! \sum_{y \in \{-1,1\}} \int I_{\{y \neq \sgn(\hat{z}^k)\}} \int p_{\Y,\hat{\Z}^k,\Z}(y, \hat{z}^k, z)        \,dz\, d\hat{z}^k\\
                &= \!\!\!\! \sum_{y\in \{-1,1\}} \iint I_{\{y \neq \sgn(\hat{z}^k)\}} p_{\Y|\Z}(y | z) p_{\Z,\hat{\Z}^k}(z, \hat{z}^k) \,dz \,d\hat{z}^k .
                \label{eq:err_rate_expression}
\end{align}
Furthermore, from the definitions of $(\Z,\hat{\Z}^k)$ and the bivariate central limit theorem, we have that
\begin{equation}
        \mat{\Z\\ \hat{\Z}^k} \xrightarrow{d} \mathcal{N}(\vec{0}, \vec{\Sigma}^k_z) = \mathcal{N}\left(\mat{0\\0}, \mat{ \Sigma^k_{11} & \Sigma^k_{12} \\
                          \Sigma^k_{21} & \Sigma^k_{22} }\right) ,
\end{equation}
where $\xrightarrow{d}$ indicates convergence in distribution.  In \cite{Z2014}, it is shown that the above matrix components are 
\ifthenelse{\boolean{ONE_COLUMN}}
{\begin{gather}
        \Sigma^k_{11} = \tau_x \sum_{n=1}^N [\var{\W_n} + \text{E}[\W_n]^2], \quad \Sigma^k_{22} = \tau_x \sum_{n=1}^N [\var{\widehat{\W}^k_n} + \text{E}[\widehat{\W}^k_n]^2], \nonumber \\
        \Sigma^k_{12} = \Sigma^k_{21} = \text{E}[\Z \hat{\Z}^k] = \tau_x \sum_{n=1}^N [\text{cov}\{\W_n, \widehat{\W}^k_n\} + \text{E}[\W_n] \text{E}[\widehat{\W}^k_n]].
        \label{eq:sigma_cross_cov}
\end{gather}}
{\begin{align}
        \Sigma^k_{11} &= \delta^{-1} \big(\var{\W_n} + \text{E}[\W_n]^2\big), \\
        \Sigma^k_{12} &= \Sigma^k_{21} = \delta^{-1} \big(\text{cov}\{\W_n, \widehat{\W}^k_n\} + \text{E}[\W_n] \text{E}[\widehat{\W}^k_n]\big),        \\
        \Sigma^k_{22} &= \delta^{-1} \big(\var{\widehat{\W}^k_n} + \text{E}[\widehat{\W}^k_n]^2\big)
        \label{eq:sigma_cross_cov}
\end{align}}
for label-to-feature ratio $\delta$.
As described earlier, the above moments can be computed using \cite[Algorithm 3]{R2011}.
The integral in \eqref{err_rate_expression} can then be computed (numerically if needed) for a given activation function $p_{\Y|\Z}$, yielding an estimate of GAMP's test-error rate at the $k^{th}$ iteration.

\begin{figure*}
        \centering
        \subfloat[][Test-Error Rate]{\scalebox{0.25}{\includegraphics*[-1.65in,0.55in][9.75in,10.1in]{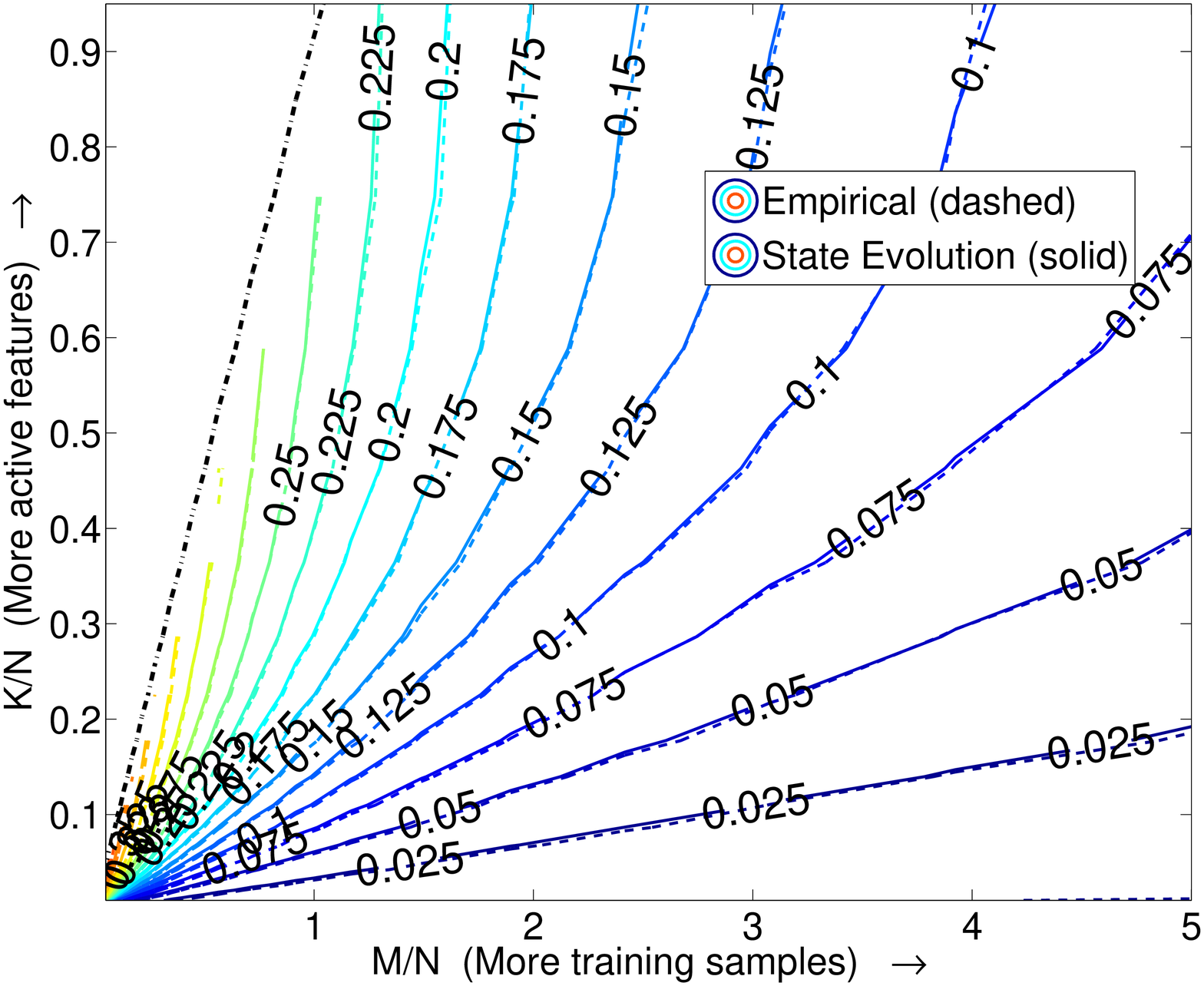}}\label{fig:state_evo_test_err}}
        \quad
        \subfloat[][Weight-Vector MSE (dB)]{\scalebox{0.25}{\includegraphics*[-1.65in,0.55in][9.75in,10.1in]{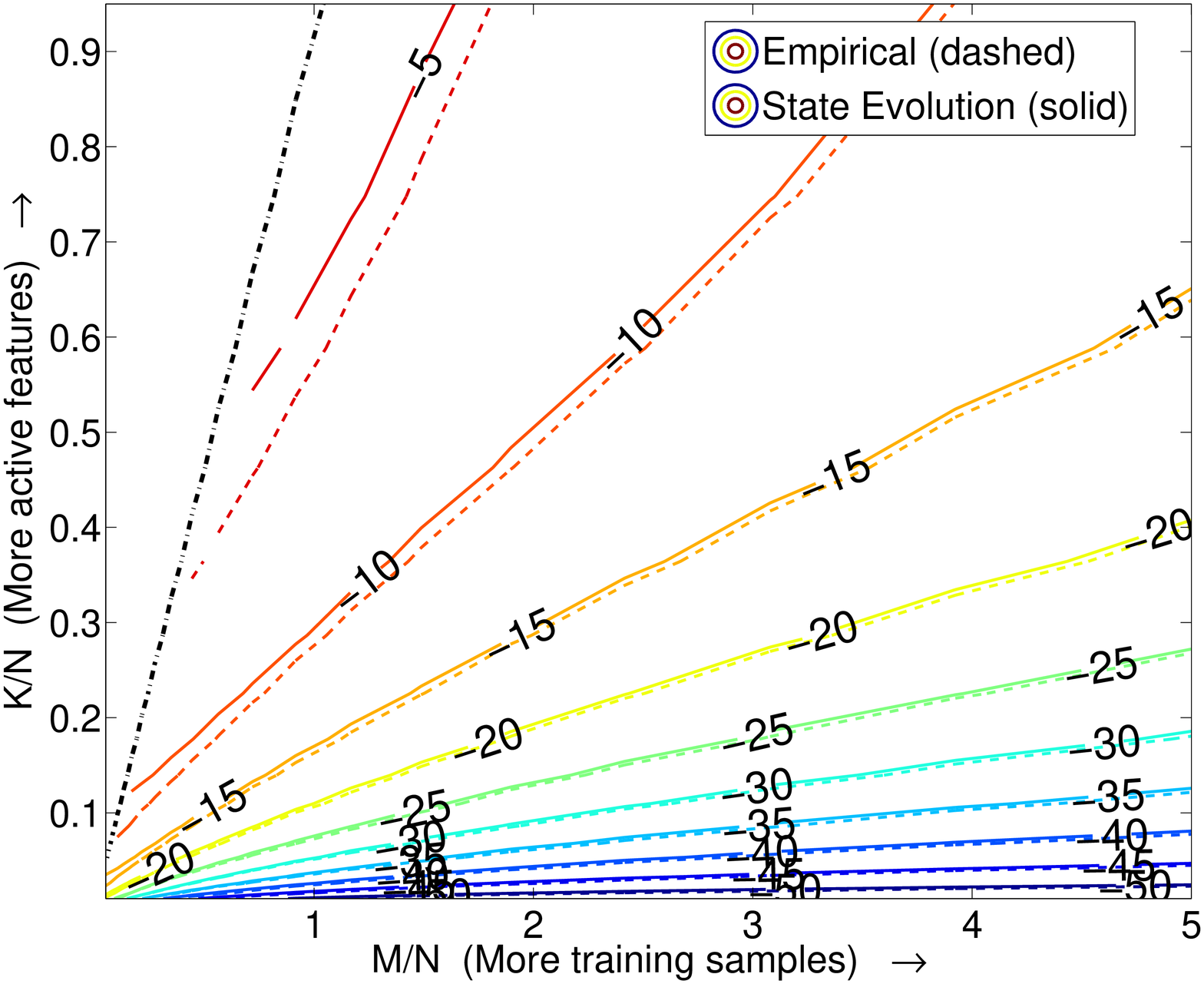}}\label{fig:state_evo_mse}}
        \caption{Test-error rate (a) and weight-vector MSE (b), versus training-to-feature ratio $M/N$ and discriminative-feature ratio $K/N$, calculated using empirical averaging (dashed) and state-evolution prediction (solid), assuming i.i.d Bernoulli-Gaussian weight vectors and a probit activation function.}
        \label{fig:state_evo_phaseplane}
\end{figure*}

To validate the accuracy of the above asymptotic analysis, we conducted a Monte-Carlo experiment with data synthetically generated in accordance with the assumed model.  In particular, 
for each of $1000$ problem realizations, a true weight vector $\vec{w}\in\Real^N$ was drawn i.i.d zero-mean Bernoulli-Gaussian and a feature matrix $\vec{X}$ was drawn i.i.d Gaussian, yielding true scores $\vec{z}=\vec{Xw}$, from which the true labels $\vec{y}$ were randomly drawn using a probit activation function $p_{\Y|\Z}$.
A GAMP weight-vector estimate $\hvec{w}^\infty$ was then computed using the training data $(\vec{y}_{1:M},\vec{X}_{1:M})$, from which the test-label estimates $\{\hat{y}_t^\infty\}_{t=M+1}^{M+T}$ with $\hat{y}_t^\infty=\sgn(\vec{x}\tran_t\hvec{w}^\infty)$ were computed and compared to the true test-labels in order to calculate the test-error rate for that realization.  \Figref{state_evo_test_err} plots the Monte-Carlo averaged empirical test-error rates (dashed) and state-evolution predicted rates (solid) as level curves over different combinations of training ratio $\frac{M}{N}$ and discriminative-feature ratio $\frac{K}{N}$, where $K=\|\vec{w}\|_0$ and $N=1024$.
Similarly, \figref{state_evo_mse} plots average empirical mean-squared error (MSE) versus state-evolution predicted MSE, where $\text{MSE}=\frac{1}{N}\text{E}\{\|\hvec{w}^\infty-\vec{w}\|_2^2\}$.

In both \figref{state_evo_test_err} and \figref{state_evo_mse}, the training-to-feature ratio $\frac{M}{N}$ increases from left to right, and the discriminative-feature ratio $\frac{K}{N}$ increases from bottom to top.  The region to the upper-left of the dash-dotted black line contains ill-posed problems (where the number of discriminative features $K$ exceeds the number of training samples $M$) for which data was not collected.  The remainders of \figref{state_evo_test_err} and \figref{state_evo_mse} show very close agreement between empirical averages and state-evolution predictions.

\section{GAMP Nonlinear Steps}
\label{sec:gamp_classification}
\Secref{gamp} gave a high-level description of how the GAMP iterations in \algref{gamp} can be applied to binary linear classification and feature selection. 
In this section, we detail the nonlinear steps used to compute $(\hat{z},\tau_z)$ and $(\hat{x},\tau_x)$ in lines \coderef{sum_prod_z}-\coderef{sum_prod_tz}, \coderef{max_sum_z}-\coderef{max_sum_tz}, \coderef{sum_prod_w}-\coderef{sum_prod_tw}, and \coderef{max_sum_w}-\coderef{max_sum_tw} of \algref{gamp}.
For sum-product GAMP, we recall that the mean and variance computations in lines \coderef{sum_prod_z}-\coderef{sum_prod_tz} and \coderef{sum_prod_w}-\coderef{sum_prod_tw} are computed based on the pdfs in \eqref{z_posterior} and \eqref{w_posterior}, respectively, and for max-sum GAMP the $\prox$ steps in \coderef{max_sum_z}-\coderef{max_sum_tz} are computed using equations \eqref{prox}-\eqref{proxderiv} and those in \coderef{max_sum_w}-\coderef{max_sum_tw} are computed similarly.

\subsection{Logistic Activation Function}
\label{sec:gamp_classification:logistic_model}
Arguably the most popular activation function for binary linear classification is the logistic sigmoid \cite[\S4.3.2]{B2006}, \cite{J1995}:
\begin{equation}
        p_{\Y|\Z}(y|z;\alpha) = \frac{1}{1 + \exp(- y \alpha z)}, ~~y\in\{-1,1\}
        \label{eq:logistic_model}
\end{equation}
where $\alpha>0$ controls the steepness of the transition.

For logistic sum-product GAMP, we propose to compute the mean and variance $(\hat{z},\tau_z)$ of the marginal posterior approximation \eqref{z_posterior} using the variational approach in \algref{variational_logistic}, whose derivation is relegated to \cite{Z2014} for reasons of space.
We note that \algref{variational_logistic} is reminiscent of the one presented in \cite[\S 10.6]{B2006}, but is more general in that it handles $\alpha\neq 1$.

For logistic max-sum GAMP, $\hat{z}$ from \eqref{prox} solves the scalar minimization problem \eqref{proxDef} with $f(u)=-\log p_{\Y|\Z}(y|u;\alpha)$ from \eqref{logistic_model}, which is convex.
To find this $\hat{z}$, we use 
bisection search to locate the root of $\frac{d}{du}[f(u)+\frac{1}{2\tau}(u-v)^2]$.
The max-sum $\tau_z$ from \eqref{proxderiv} can then be computed in closed form using $\hat{z}$ and $f''(\cdot)$ via \eqref{proxderiv2}.
Note that, unlike the classical ML-based approach to logistic regression (e.g., \cite[\S4.3.3]{B2006}), GAMP performs only scalar minimizations and thus does not need to construct or invert a Hessian matrix.

\begin{Algorithm}{2.95in}
        \caption{A Variational Approach to Logistic Activation Functions for Sum-Product GAMP}
        \label{alg:variational_logistic}
        \singlespacing
        \begin{algorithmic}[1]
                \REQUIRE{Class label $y \in \{-1,1\}$, logistic scale $\alpha$, and GAMP-computed parameters $\hat{p}$ and $\tau_p$ (see \eqref{z_posterior})}
                \ENSURE{$\xi \leftarrow \smash{\sqrt{\tau_p + |\hat{p}|^2}}$}
                \REPEAT
                        \STATE $\sigma \leftarrow \big(1 + \exp(-\alpha \xi)\big)^{-1}$
                        \STATE $\lambda \leftarrow \tfrac{\alpha}{2 \xi}(\sigma - \tfrac{1}{2})$
                        \STATE $\tau_z \leftarrow \tau_p (1 + 2 \tau_p \lambda)^{-1}$
                        \STATE $\hat{z} \leftarrow \tau_z (\hat{p}/\tau_p + \alpha y / 2)$
                        \STATE $\xi \leftarrow \sqrt{\tau_z + |\hat{z}|^2}$
                \UNTIL{Terminated}
                \RETURN{$\hat{z}$, $\tau_z$}
        \end{algorithmic}
\end{Algorithm}

\subsection{Probit Activation Function}
\label{sec:gamp_classification:probit_model}
Another popular activation function is the probit \cite[\S4.3.5]{B2006}:
\begin{eqnarray}
        p_{\Y|\Z}(1 | z;v) &=& \int_{-\infty}^{z} \mathcal{N}(\tau; 0, v) d\tau 
                = \Phi\Big(\frac{z}{\sqrt{v}}\Big)
        \label{eq:probit_model}
\end{eqnarray}
where $p_{\Y|\Z}(-1 | z) = 1 - p_{\Y|\Z}(1 | z) = \Phi(-\tfrac{z}{\sqrt{v}})$ and where $v > 0$ controls the steepness of the sigmoid. 

Unlike the logistic case, the probit case leads to closed-form sum-product GAMP computations.  In particular, the density \eqref{z_posterior} corresponds to the posterior pdf of a random variable $\Z$ with prior $\mathcal{N}(\hat{p},\tau_p)$ from an observation $\Y=y$ measured under the likelihood model \eqref{probit_model}.  A derivation in \cite[\S3.9]{RW2006} provides the necessary expressions for these moments when $y \!=\! 1$, and a similar exercise tackles the $y \!=\! -1$ case.  For completeness, the sum-product computations are summarized in \tabref{probit_updates}.  Max-sum GAMP computation of $(\hat{z},\tau_z)$ can be performed using a bisection search akin to that described in  \secref{gamp_classification:logistic_model}.

\begin{table}[t]
        \centering
        \renewcommand{\arraystretch}{1.45}
        \begin{tabular}{c|c}
                Quantity & Value        \\
                \hline
                $c$ & $\displaystyle \frac{\hat{p}}{\sqrt{v + \tau_p}}$ \\ 
                $\hat{z}$ & $\hat{p} + \displaystyle \frac{y\tau_p \phi({c})}{\Phi(y{c}) \sqrt{v + \tau_p}}$ \\
                $\tau_z$ & $\tau_p - \displaystyle \frac{\tau_p^2 \phi({c})}{\Phi(y{c}) (v + \tau_p)}\left(y{c} + \displaystyle \frac{\phi({c})}{\Phi(c)}\right)$ \\ 
        \end{tabular}
        \caption{Sum-product GAMP computations for probit activation function.}
        \label{tab:probit_updates}
\end{table}

\subsection{Hinge-Loss Activation Function}
\label{sec:gamp_classification:hinge_loss_model}
The \emph{hinge loss} $f_{\Z_m}\!(z) \triangleq \max(0, 1 - y_m z)$ is commonly used in the \emph{support vector machine} (SVM) approach to maximum-margin classification \cite[\S7.1]{B2006}, i.e., 
\begin{eqnarray}
        \hvec{w} = \argmin_{\vec{w}} \sum_{m=1}^M \!f_{\Z_m}\!(\vec{x}_m\tran \vec{w}) + \lambda \|\vec{w}\|_2^2
        \label{eq:svm_optimization}
\end{eqnarray}
or variations where $\|\vec{w}\|_2^2$ is replaced with a sparsity-inducing alternative like $\|\vec{w}\|_1$ \cite{BBEBS2003}.
Recalling \secref{gamp:max_sum}, this loss leads to the activation function 
\begin{eqnarray}
        p_{\Y_m|\Z_m}(y_m|z) &\propto& \exp\big(-\max(0, 1 - y_m z)\big).
        \label{eq:hinge_prior}
\end{eqnarray}

For hinge-loss sum-product GAMP, the mean and variance $(\hat{z},\tau_z)$ of \eqref{z_posterior} can be computed in closed form using the procedure described in \appref{sum_product_hinge}, and summarized in \tabref{hinge_updates}.
Meanwhile, for max-sum GAMP, the proximal steps \eqref{prox}-\eqref{proxderiv} can be efficiently computed using bisection search, as in the logistic and probit cases.  

\begin{table}[t]
        \centering
        \renewcommand{\arraystretch}{1.45}
        \begin{tabular}{c|c}
                Quantity & Value        \\
                \hline
                $\hat{z}$ & $\displaystyle (1 + \gamma_y)^{-1} \ubar{\mu}_y + (1 + \gamma_y^{-1})^{-1} \bar{\mu}_y$ \\
                $\tau_z$ & $\displaystyle (1 \!+\! \gamma_y)^{-1} (\ubar{v}_y \!+\! \ubar{\mu}_y^2) \!+\! (1 \!+\! \gamma_y^{-1})^{-1} (\bar{v}_y \!+\! \bar{\mu}_y^2) \!-\! \hat{z}^2$ \\ 
        \end{tabular}
        \caption{Sum-product GAMP computations for the hinge-loss activation function.  See \appref{sum_product_hinge} for definitions of $\gamma_y$, $\underline{\mu}_y$, $\bar{\mu}_y$, $\underline{v}_y$, $\bar{v}_y$.
        }
        \label{tab:hinge_updates}
\end{table}

\subsection{A Method to Robustify Activation Functions}
\label{sec:gamp_classification:robust}
In some applications, a fraction $\gamma\in(0,1)$ of the training labels are known\footnote{A method to learn an unknown $\gamma$ will be proposed in \secref{parameter_tuning}.} to be corrupted, or at least highly atypical under a given activation model $p_{\Y|\Z}^*(y|z)$.
As a robust alternative to $p_{\Y|\Z}^*(y|z)$, Opper and Winther \cite{OW2000} proposed to use
\begin{eqnarray}
        p_{\Y|\Z}(y|z;\gamma) &=& (1 - \gamma) p_{\Y|\Z}^*(y|z) + \gamma p_{\Y|\Z}^*(-y|z)      \\
                &=& \gamma + (1 - 2 \gamma) p_{\Y|\Z}^*(y|z).
                \label{eq:robust_model}
\end{eqnarray}
We now describe how the GAMP nonlinear steps for an arbitrary $p_{\Y|\Z}^*$ can be used to compute the GAMP nonlinear steps for a robust $p_{\Y|\Z}$ of the form in \eqref{robust_model}.

In the sum-product case, knowledge of the non-robust quantities 
$\hat{z}^* \triangleq \tfrac{1}{C^*_{y}} \int_z z \, p_{\Y|\Z}^*(y|z) \mathcal{N}(z;\hat{p},\tau_p)$, 
$\tau_z^* \triangleq \tfrac{1}{C^*_{y}} \int_z (z - \hat{z}^*)^2\, p_{\Y|\Z}^*(y|z)\mathcal{N}(z;\hat{p},\tau_p)$,
and 
$C^*_{y} \triangleq \int_z p_{\Y|\Z}^*(y|z) \mathcal{N}(z;\hat{p},\tau_p)$ 
is sufficient for computing the robust sum-product quantities $(\hat{z},\tau_z)$, as summarized in \tabref{robust_updates}.  (See \cite{Z2014} for details.)  

In the max-sum case, computing $\hat{z}$ in \eqref{prox} involves solving the scalar minimization problem in \eqref{proxDef} with $f(u)=-\log p_{\Y|\Z}(y|u;\gamma)=-\log[\gamma + (1-2\gamma) p_{\Y|\Z}^*(y|u)]$.
As before, we use a bisection search to find $\hat{z}$ and then we use $f''(\hat{z})$ to compute $\tau_z$ via \eqref{proxderiv2}. 

\begin{table}[t]
        \centering
        \renewcommand{\arraystretch}{1.45}
        \begin{tabular}{c|c}
                Quantity & Value        \\
                \hline
                $C_{y}$ & $\displaystyle \frac{\gamma}{\gamma + (1 - 2 \gamma) C^*_{y}}$ \\ 
                $\hat{z}$ & $C_{y} \hat{p} + (1-C_{y}) \hat{z}^*$ \\
                $\tau_z$ & $C_{y} (\tau_p + \hat{p}^2) + (1-C_{y}) (\tau_z^* + (\hat{z}^*)^2) - \hat{z}^2$ \\ 
        \end{tabular}
        \caption{Sum-product GAMP computations for a robustified activation function.
                 See text for definitions of $C^*_{y}$, $\hat{z}^*$, and $\tau_z^*$.}
        \label{tab:robust_updates}
\end{table}

\subsection{Weight Vector Priors}
\label{sec:gamp_classification:weights}

We now discuss the nonlinear steps used to compute $(\hat{w},\tau_w)$, i.e., lines \coderef{sum_prod_w}-\coderef{sum_prod_tw} and \coderef{max_sum_w}-\coderef{max_sum_tw} of \algref{gamp}.
These steps are, in fact, identical to those used to compute $(\hat{z},\tau_z)$ except that the prior $p_{\W_n}(\cdot)$ is used in place of the activation function $p_{\Y_m|\Z_m}(y_m|\cdot)$.
For linear classification and feature selection in the $N\gg M$ regime, it is customary to choose a prior $p_{\W_n}(\cdot)$ that leads to sparse (or approximately sparse) weight vectors $\vec{w}$, as discussed below.

For sum-product GAMP, this can be accomplished by choosing a Bernoulli-$\tilde{p}$ prior, i.e.,
\begin{equation}
  p_{\W_n}\!(w) = (1-\pi_n) \delta(w) + \pi_n \tilde{p}_{\W_n}\!(w), \label{eq:Bernoulli}
\end{equation}
where $\delta(\cdot)$ is the Dirac delta function, $\pi_n\in[0,1]$ is the prior\footnote{In \secref{parameter_tuning} we describe how a common $\pi=\pi_n~\forall n$ can be learned.
} 
probability that $\W_n\!=\!0$, and $\tilde{p}_{\W_n}\!(\cdot)$ is the pdf of a non-zero $\W_n$.
While Bernoulli-Gaussian \cite{S2010a} and Bernoulli-Gaussian-mixture \cite{VS2013} are common choices, \secref{numerical_study} suggests that Bernoulli-Laplacian also performs well. 

In the max-sum case, the GAMP nonlinear outputs $(\hat{w},\tau_w)$ are computed via 
\begin{align}
\hat{w}&=\prox_{\tau_r f_{\W_n}}\!(\hat{r}) \\
\tau_w &=\tau_r \prox'_{\tau_r f_{\W_n}}\!(\hat{r})
\end{align}
for a suitably chosen regularizer $f_{\W_n}\!(w)$.
Common examples include 
$f_{\W_n}\!(w)=\lambda_1 |w|$ for $\ell_1$ regularization \cite{DMM2009},
$f_{\W_n}\!(w)=\lambda_2 w^2$ for $\ell_2$ regularization \cite{R2011}, and
$f_{\W_n}\!(w)=\lambda_1 |w| + \lambda_2 w^2$ for the ``elastic net'' \cite{ZH2005}.
As described in \secref{gamp:max_sum}, any regularizer $f_{\W_n}$ can be interpreted as a (possibly improper) prior pdf $p_{\W_n}\!(w)\propto \exp(-f_{\W_n}\!(w))$.
Thus, $\ell_1$ regularization corresponds to a Laplacian prior, $\ell_2$ to a Gaussian prior, and the elastic net to a product of Laplacian and Gaussian pdfs.  

In \tabref{gamp_priors}, we give the sum-product and max-sum computations for the prior corresponding to the elastic net, which includes both Laplacian (i.e., $\ell_1$) and Gaussian (i.e., $\ell_2$) as special cases; a full derivation can be found in \cite{Z2014}.
For the Bernoulli-Laplacian case, these results can be combined with the Bernoulli-$\tilde{p}$ extension in \tabref{gamp_priors}.

\begin{table}[t]
        \centering
        \renewcommand{\arraystretch}{1.45}
        \begin{tabular}{cc|c}
                & Quantity & Value      \\
                \hline
                \multirow{2}{*}{\rotatebox{90}{SPG\,}} & $\hat{w}$ & $\displaystyle \big(\ubar{C} \ubar{\mu} + \bar{C} \bar{\mu}\big) / \big(\ubar{C} + \bar{C}\big)$ \\
                & $\tau_w$ & $\displaystyle \big(\ubar{C} (\ubar{v} + \ubar{\mu}^2) + \bar{C} (\bar{v} + \bar{\mu}^2)\big) / \big(\ubar{C} \!+\! \bar{C}\big) - \hat{w}^2$ \\ 
                \hline
                \multirow{2}{*}{\rotatebox{90}{MSG\,}} & $\hat{w}$ & $\displaystyle \sgn(\sigma \ddot{r}) \max(|\sigma \ddot{r}| - \lambda_1 \sigma^2, 0)$ \\
                & $\tau_w$ & $\displaystyle \sigma^2 \cdot \text{I}_{\{\hat{w} \neq 0\}}$ \\
        \end{tabular}
        \caption{Sum-product GAMP (SPG) and max-sum GAMP (MSG) computations for the elastic-net regularizer $f_{\W_n}\!(w)=\lambda_1 |w| + \lambda_2 w^2$, which includes $\ell_1$ or Laplacian-prior (via $\lambda_2\!=\!0$) and $\ell_2$ or Gaussian-prior (via $\lambda_1\!=\!0$) as special cases.  See \tabref{elastic_defns} for definitions of $\underline{C}$, $\bar{C}$, $\underline{\mu}$, $\bar{\mu}$, etc.}
        \label{tab:elastic_updates}
\end{table}

\begin{table}[t]
        \centering
        \renewcommand{\arraystretch}{1.45}
        \begin{tabular}{c|c}
                \hline
                 $\sigma \triangleq \sqrt{\tau_r / (2 \lambda_2 \tau_r + 1)}$ & $\ddot{r} \triangleq \hat{r} / (\sigma (2 \lambda_2 \tau_r + 1))$       \\
                $\ubar{r} \triangleq \ddot{r} + \lambda_1 \sigma$ & $\bar{r} \triangleq \ddot{r} - \lambda_1 \sigma$    \\
                $\ubar{C} \triangleq \frac{\lambda_1}{2} \exp \big(\frac{\ubar{r}^2 - \ddot{r}^2}{2}\big) \Phi (\textendash \ubar{r})$ & $\bar{C} \triangleq \frac{\lambda_1}{2} \exp \big(\frac{\bar{r}^2 - \ddot{r}^2}{2}\big) \Phi (\bar{r})$        \\
                $\ubar{\mu} \triangleq \sigma \ubar{r} - \sigma \phi (\textendash \ubar{r}) / \Phi (\textendash \ubar{r})$ & $\bar{\mu} \triangleq \sigma \bar{r} + \sigma \phi (\bar{r}) / \Phi (\bar{r})$     \\
                $\ubar{v} \triangleq \sigma^2 \left[1 \!-\! \frac{\phi(\ubar{r})}{\Phi(\ubar{r})} \!\! \left(\frac{\phi(\ubar{r})}{\Phi(\ubar{r})} \!-\! \ubar{r} \right) \! \right]$ & $\bar{v} \triangleq \sigma^2 \left[1 \!-\! \frac{\phi(\bar{r})}{\Phi(\bar{r})} \!\! \left(\frac{\phi(\bar{r})}{\Phi(\bar{r})} \!+\! \bar{r} \right) \! \right]$ \\
                \hline
        \end{tabular}
        \caption{Definitions of elastic-net quantities used in \tabref{elastic_updates}.}
        \label{tab:elastic_defns}
\end{table}

\subsection{The GAMPmatlab Software Suite}
\label{sec:gamp_classification:software}
The GAMP iterations from \algref{gamp}, including the nonlinear steps discussed in this section, have been implemented in the open-source ``GAMPmatlab'' software suite.\footnote{The latest source code can be obtained through the GAMPmatlab SourceForge Subversion repository at \url{http://sourceforge.net/projects/gampmatlab/}.}
For convenience, the existing activation-function implementations are summarized in \tabref{gamp_likelihoods} and relevant weight-prior implementations appear in \tabref{gamp_priors}.

\begin{table}[t]
        \centering
        \begin{tabular}{c|c|c@{~~}c}
                 Name & $p_{\Y|\Z}(y|z)$ Description & Sum-& Max-\\[-1mm]
                      & & Product & Sum \\
                 \hline
                 Logistic & $\propto (1 + \exp(-\alpha y z))^{-1}$ & VI & RF    \\
                 Probit & $\Phi\big(\tfrac{yz}{v}\big)$ & CF & RF       \\
                 Hinge Loss & $\propto \exp(-\max(0, 1 - yz))$ & CF & RF        \\
                 \hline
                 Robust-$p^*$ & $\gamma + (1 - 2\gamma) p_{\Y|\Z}^*(y|z)$ & CF & RF
        \end{tabular}
        \caption{Activity-functions and their GAMPmatlab sum-product and max-sum implementation method: CF = closed form, VI = variational inference, RF = root-finding.}
        \label{tab:gamp_likelihoods}
\end{table}

\begin{table}[t]
        \centering
        \begin{tabular}{c|c|c@{~~}c}
                 Name & $p_{\W_n}\!(w)$ Description        & Sum-    & Max-\\[-1mm]
                      & & Product & Sum \\
                 \hline
                 Gaussian & $\mathcal{N}(w; \mu, \sigma^2)$ & CF & CF   \\
                 GM & $\sum_l \omega_l \mathcal{N}(w; \mu_l, \sigma_l^2)$ & CF & NI     \\
                 Laplacian & $\propto \exp(-\lambda |w|)$ & CF & CF     \\
                 Elastic Net & $\propto \exp(-\lambda_1 |w| - \lambda_2 w^2)$ & CF & CF \\
                 \hline
                 Bernoulli-$\tilde{p}$ & $(1-\pi_n) \delta(w) + \pi_n \tilde{p}_{\W_n}\!(w)$ & CF & NA
        \end{tabular}
        \caption{Weight-coefficient priors and their GAMPmatlab sum-product and max-sum implementation method: CF = closed form, NI = not implemented, NA = not applicable.}
        \label{tab:gamp_priors}
\end{table}

\section{Online Parameter Tuning}
\label{sec:parameter_tuning}
The activation functions and weight-vector priors described in \secref{gamp_classification} depend on modeling parameters that, in practice, must be tuned.
For example, the logistic sigmoid \eqref{logistic_model} depends on $\alpha$; the probit depends on $v$; $\ell_1$ regularization depends on $\lambda$; and the Bernoulli-Gaussian-mixture prior depends on $\pi$ and $\{\omega_l,\mu_l,\sigma^2_l\}_{l=1}^L$, where $\omega_l$ parameterizes the weight, $\mu_l$ the mean, and $\sigma^2_l$ the variance of the $l$th mixture component.
Although cross-validation (CV) is the customary approach to tuning parameters such as these, it suffers from two major drawbacks:
First, it can be very computationally costly, since each parameter must be tested over a grid of hypothesized values and over multiple data folds.
For example, $K$-fold cross-validation tuning of $P$ parameters using $G$ hypothesized values of each requires the training and evaluation of $KG^P$ classifiers.
Second, leaving out a portion of the training data for CV can degrade classification performance, especially in the example-starved regime where $M\ll N$ (see, e.g., \cite{NMTM2000}).

As an alternative to CV, we consider \emph{online learning} of the unknown model parameters $\vec{\theta}$ using the methodology from \cite{VS2013,Vila:TSP:14}.
Here, the goal is to compute the maximum-likelihood estimate $\hvec{\theta}_{\textsf{ML}}=\argmax_{\vec{\theta}}p_{\vY}(\vec{y};\vec{\theta})$, where our data model implies a likelihood function of the form
\begin{equation}
p_{\vY}(\vec{y};\vec{\theta})
= \int_{\vec{w}} \prod_m p_{\Y_m|\Z_m}(y_m|\vec{x}\tran\vec{w};\vec{\theta}) \prod_n p_{\W_n}\!(w_n;\vec{\theta}) .
\label{eq:like}
\end{equation}
Because it is computationally infeasible to evaluate and/or maximize \eqref{like} directly, we apply the expectation-maximization (EM) algorithm \cite{DLR1977}.
For EM, we treat $\vW$ as the ``hidden'' data, giving the iteration-$j$ EM update
\begin{align}
\vec{\theta}^{j} 
&= \argmax_{\vec{\theta}} \text{E}_{\vW|\vY}\big\{\log \, p_{\vY,\vW}(\vec{y}, \vW; \vec{\theta}) \,\big|\, \vec{y}; \vec{\theta}^{j-1}\big\} \\
&= \argmax_{\vec{\theta}} 
\sum_m \text{E}_{\Z_m|\vY} \big\{ \log p_{\Y_m|\Z_m}(y_m|\Z_m; \vec{\theta}) \,\big|\, \vec{y}; \vec{\theta}^{j-1}\big\} 
\nonumber\\&\quad
+\sum_n \text{E}_{\W_n|\vY} \big\{ \log p_{\W_n}(\W_n; \vec{\theta}) \,\big|\, \vec{y}; \vec{\theta}^{j-1}\big\} .  \label{eq:generic_em}
\end{align}
Furthermore, to evaluate the conditional expectations in \eqref{generic_em}, GAMP's posterior approximations from \eqref{z_posterior}-\eqref{w_posterior} are used.
It was shown in \cite{KRFU2012} that, in the large-system limit, the estimates generated by this procedure are asymptotically consistent (as $j\rightarrow\infty$ and under certain identifiability conditions).
Moreover, it was shown in \cite{VS2013,Vila:TSP:14} that, for various priors and likelihoods of interest in compressive sensing (e.g., AWGN likelihood, Bernoulli-Gaussian-Mixture priors, $\ell_1$ regularization), the quantities needed from the expectation in \eqref{generic_em} are implicitly computed by GAMP, making this approach computationally attractive.
However, because this EM procedure runs GAMP several times, once for each EM iteration (although not necessarily to convergence), the total runtime may be increased relative to that of GAMP without EM.

In this work, we propose EM-based learning of the activation-function parameters, i.e., $\alpha$ in the logistic model \eqref{logistic_model}, $v$ in the probit model \eqref{probit_model}, and $\gamma$ in the robust model \eqref{robust_model}.
Starting with $\alpha$, we find that a closed-form expression for the value maximizing \eqref{generic_em} remains out of reach, due to the form of the logistic model \eqref{logistic_model}. 
So, we apply the same variational lower bound used for \algref{variational_logistic}, and find that the lower-bound maximizing value of $\alpha$ obeys (see \cite{Z2014})
\begin{equation}
    0 = \sum_m \tfrac{1}{2}(\hat{z}_m y_m - \xi_m) + \frac{\xi_m}{1 + \exp(\alpha \xi_m)} 
    \label{eq:alpha_em_upd} ,
\end{equation}
where $\xi_m$ is the variational parameter being used to optimize the lower-bound and $\hat{z}_m \approx \text{E}\{\Z_m|\vY=\vec{y}\}$ is output by \algref{variational_logistic}.
We then solve for $\alpha$ using Newton's method.

To tune the probit parameter, $v$, we zero the derivative of \eqref{generic_em} w.r.t $v$ to obtain
\begin{eqnarray}
        0 
        &=& \sum_m \text{E}_{\Z_m|\vY} \Big\{ \tfrac{\partial}{\partial v} \log p_{\Y_m|\Z_m}(y_m|\Z_m; v^j) \,\Big|\, \vec{y}; v^{j-1}\Big\}   \\
        &=& \sum_m \text{E}_{\Z_m|\vY} \Big\{ \tfrac{-\ddot{c}_m\!(v^j)}{v^j} \phi(\ddot{c}_m\!(v^j)) \Phi(\ddot{c}_m\!(v^j))^{-1} \Big| \vec{y}; v^{j-1}\Big\},
        \qquad
        \label{eq:probit_em_upd}
\end{eqnarray}
where $\ddot{c}_m(v) \triangleq (y_m z_m) / v$.  
We then numerically evaluate the expectation and apply an iterative root-finding procedure to find the EM update $v^j$ that solves \eqref{probit_em_upd}.

To learn $\gamma$, we include the corruption indicators $\vec{\beta}\!\in\!\{0,1\}^M$ in the EM-algorithm's hidden data (i.e., $\beta_m\!=\!0$ indicates that $y_m$ was corrupt and $\beta_m\!=\!1$ that it was not), where an i.i.d assumption on the corruption mechanism implies the prior $p(\vec{\beta};\gamma) = \prod_{m=1}^M \gamma^{1-\beta_m} (1 - \gamma)^{\beta_m}$.  
In this case, it can be shown \cite{Z2014} that the update of the $\gamma$ parameter reduces to 
\begin{align}
       \gamma^{j} 
       &= \argmax_{\gamma\in[0,1]} \sum_{m=1}^M \text{E}_{\beta_m | \vY}\big[\log \, p(\beta_m; \gamma) \,\big|\, \vec{y}; \vec{\theta}^{j-1}\big]
       \label{eq:gamma_em_update}
       \\
       &= \frac{1}{M} \sum_{m=1}^M p(\beta_m \!=\! 0 \,|\, \vec{y};\vec{\theta}^{j-1}),
       \label{eq:gamma_em_update2}
\end{align}
where \eqref{gamma_em_update2} leveraged $\text{E}[\beta_m | \vec{y}; \vec{\theta}^{j-1}] = 1-p(\beta_m \!=\! 0 | \vec{y};\vec{\theta}^{j-1})$.
Moreover, $p(\beta_m \!=\! 0 | \vec{y};\vec{\theta}^{j-1})$ is easily computed using quantities returned by sum-product GAMP.


\section{Numerical Study}
\label{sec:numerical_study}
In this section we describe several synthetic and real-\textb{world} classification problems to which GAMP was applied.  
Experiments were conducted on a workstation running Red Hat Enterprise Linux (r2.4), with an Intel Core i7-2600 CPU (3.4 GHz, 8MB cache) and 8GB DDR3 RAM. 

\subsection{Synthetic Classification in the $N\gg M$ Regime}
\label{sec:numerical_study:em}
\textb{We first examine a synthetic problem where the number of features, $N$, greatly exceeds the number of training examples, $M$. 
As discussed in the Introduction, it is possible to perform accurate classification when $N\!\gg\! M$ if the number of discriminatory features $K$ is sufficiently small. 
In this experiment, we consider $N\!=\!30\,000$, $M\!=\!300$, and $K\!\in\!\{5,\dots,30\}$, where the range on $K$ is chosen based on the following information-theoretic argument: 
$M$ training labels bring $\log_2 M$ bits of information, whereas at least $K \log_2 (N/K) \!\leq\! \log_2 {N \choose K}$ bits of information are needed to determine the $N$-length $K$-sparse Bayes weight vector, assuming that we have no prior knowledge of its support, which takes on ${N \choose K}$ possibilities.
With $N\!=\!30\,000$ and $M\!=\!300$, it turns out that $K=31$ is the largest value of $K\leq N$ such that $M\geq K \log_2 (N/K)$.}

\textb{Our experiment was of a Monte-Carlo form. 
In each trial, we constructed a random $K$-sparse Bayes weight vector $\vec{w}$ with a support drawn uniformly at random and with non-zero-coefficient amplitudes drawn uniformly in $\{-1,1\}$.
We used $\pm 1$ amplitudes to eliminate the potential ambiguity about whether a given non-zero coefficient was effectively non-zero, since, e.g., Gaussian-distributed amplitudes can be arbitrarily close to zero. 
We then constructed a balanced set of training labels $y_m\in\{-1,1\}$ (i.e., exactly $M/2$ labels were positive) and we drew $M$ i.i.d random feature vectors $\vec{x}_m$ from the class-conditional generative distribution $\vec{x}_m|y_m \sim \mc{N}(y_m \vec{w},v\vec{I})$.}

\textb{\Figref{em_sparsity_accuracy_test} shows both the average test error rate and the average estimated sparsity $\hat{K}$ for cross-validation tuned ``OneBitCS'' from \cite{PV2013},\footnote{\textb{For cross-validation of OneBitCS, we used $2$ folds and searched over all sparsities in a radius of $10$ from the true sparsity $K$.}} and for EM-tuned sum-product GAMP classifiers based on the Bernoulli-Gaussian (BG) prior and activation functions including hinge loss (HL), probit (PR), and logistic (LR).
The average was computed over $50$ Monte-Carlo trials, where in each trial the expected error probability of the designed classifier $\hvec{w}$ was computed in closed form as $\Phi(-\vec{w}\tran\hvec{w}/\sqrt{v \|\hvec{w}\|^2})$.
The figure shows all algorithms under test performing relatively close to the Bayes error rate, and for small $K$ it shows BG-LR and BG-PR GAMP performing extremely close to the Bayes error rate.
Comparing the classifiers, we see that GAMP's BG-LR performs the best, which is not surprising since the logistic activation function is statistically matched to data model in this experiment \cite{J1995}.
Meanwhile, GAMP's BG-PR classifier performed the second best, and the two remaining classifiers (GAMP's BG-HL and OneBitCS) performed only slightly worse.}

\textb{\Figref{em_sparsity_accuracy_test} also shows the sparsities estimated by cross-validation in the case of OneBitCS and by the EM-tuning in the case of GAMP.
Since the weights returned by sum-product BG-GAMP are non-zero with probability one, the estimated sparsity is defined as the number of coefficients with posterior support probability $p(w_n \!\neq\! 0 | \vec{y})$ exceeding $1/2$.
The figure shows that all algorithms under test returned accurate estimates of the true sparsity $K$.
For small values of $K$, the estimates returned by BG-LR and BG-PR GAMP were extremely accurate while those for OneBitCS and BG-HL GAMP slightly overestimated the sparsity. 
Meanwhile, for large values of $K$, all algorithms underestimated the sparsity by about $15$\%.}

\begin{figure}
        \begin{center}
                \ifthenelse{\boolean{ONE_COLUMN}}
                {\includegraphics[width=3.5in]{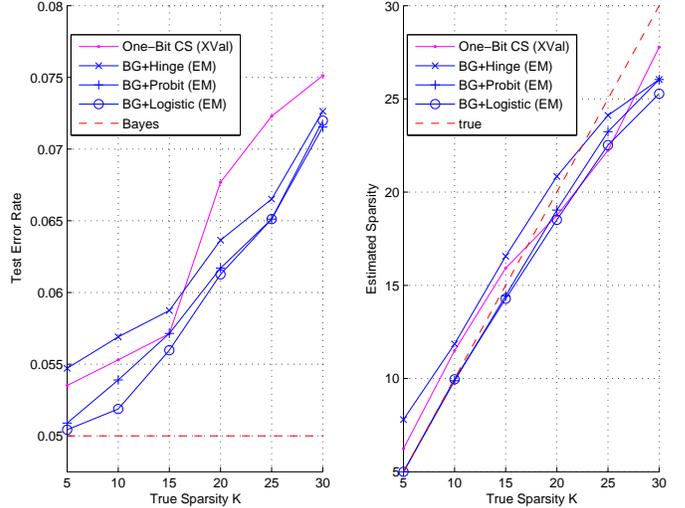}}
                {\includegraphics[width=3.5in]{Figures/em_sparsity_accuracy_test.eps}}
                \caption{\textb{Test error rate and estimated sparsity $\hat{K}$ for cross-validation-tuned OneBitCS, and for EM-tuned sum-product GAMP classifiers based on the Bernoulli-Gaussian (BG) prior and the hinge ($\times$), probit ($+$), and logistic ($\circ$) activation functions, as a function of the true sparsity $K$.  Here, $N=30\,000$, $M=300$, and Bayes error rate was $\eps_{\text{B}} = 0.05$.}}
                \label{fig:em_sparsity_accuracy_test}
        \end{center}
\end{figure}

\subsection{Text Classification and Adaptive Learning}
\label{sec:numerical_study:text}
We \textb{next} consider a binary text classification problem based on the Reuter's Corpus Volume I (RCV1) dataset \cite{LYRL2004}.  
As in \cite{YCHL2010,LWK2007}, newswire article topic codes \texttt{CCAT} and \texttt{ECAT} were combined to form the positive class while \texttt{GCAT} and \texttt{MCAT} were combined to form constitute the negative class.\footnote{Data was taken from \burl{http://www.csie.ntu.edu.tw/~cjlin/libsvmtools/datasets/binary.html}.} 
Although the original dataset consisted of $20\,242$ balanced training examples of $N \!=\! 47\,236$ features, with $677\,399$ examples reserved for testing, we followed the approach in \cite{LWK2007,YCHL2010} and swapped training and testing sets in order to test computational efficiency on a large training dataset (and thus $M \!=\! 677\,399$).  
As in \cite{YCHL2010}, we constructed feature vectors as cosine-normalized logarithmic transformations of the TF-IDF (term frequency -- inverse document frequency) data vectors.  
We note that the resulting features are very sparse; only $0.16\%$ of the entries in $\vec{X}$ are non-zero.  
Finally, we trained linear classifiers (i.e., weight vectors) using four GAMP-based methods and \textb{four} existing state-of-the-art methods: 
TFOCS \cite{BCG2011} in L1-LR mode, CDN \cite{YCHL2010}, TRON \cite{LM1999}\textb{, and OneBitCS \cite{PV2013}}. 
In doing so, for EM learning we used $5$ EM iterations, and for cross-validation we used $2$ folds and a logarithmically spaced grid of size $10$.\footnote{\textb{For OneBitCS, the cross-validation grid included sparsity rates between $0.1$\% and $15$\%.}}  

\tabref{rcv1_performance_flipped} summarizes the performance achieved by the resulting classifiers, including the test-set classification accuracy, weight-vector density (i.e., the fraction of non-zero weights), and two runtimes: the \emph{total} runtime needed to train the classifier, which includes EM- or cross-validation-based parameter tuning, and the \emph{post-tuning} runtime.  
Although it is customary to report only the latter, we feel that the former better captures the true computational cost of classifier design.
\textb{We note that, in the case of spGAMP, the total and post-tuning runtime are identical because EM tuning was performed once per GAMP iteration. 
In contrast, for msGAMP, we ran many GAMP iterations per EM iteration, and hence the total runtime (which avoids EM iterations) is much longer.
We also note that the post-tuning runtime of OneBitCS is extremely fast because of a computational trick that we learned via personal communication with an author, Yaniv Plan: Given signed labels $y_m\in\{-1,1\}$ and a sparsity estimate $\hat{K}$, the OneBitCS weight vector $\hvec{w}$ can be computed from the training pair $(\vec{X},\vec{y})$ via $\hvec{w}=\text{thresh}_{\hat{K}}(\vec{X}\tran\vec{y})$, where $\text{thresh}_{\hat{K}}(\cdot)$ is the mapping from $\Real^N\rightarrow\Real^N$ that preserves the input components with the largest $\hat{K}$ magnitudes and zeros the remainder.}

\begin{table}[t]
\color{blue}
        \centering
        \begin{tabular}{@{}c@{~}c|c@{~}c@{~}c@{}}
                Classifier & Tuning & Accuracy & Runtime (s) & Density  \\
                \hline
                spGAMP:\,BG-PR & EM & 97.4\% & {\bf 105} / 105 & 8.6\%        \\
                spGAMP:\,BG-HL & EM & 97.3\% & 134 / 134 & 8.9\%     \\
                msGAMP:\,L1-LR & EM & 97.6\% & 684 / 123 & 9.8\%        \\
                msGAMP:\,L1-LR & xval & 97.6\% & 3068 / 278 & 19.6\%    \\
                
                \hline

                CDN \cite{YCHL2010} & xval & \bf 97.7\% & 1298 / 112 & 10.9\% \\
                TRON \cite{LM1999} & xval & \bf 97.7\% & 1682 / 133 & 10.8\% \\
                
                TFOCS \cite{BCG2011} & xval & 97.6\% & 1086 / 94 & 19.2\% \\
                OneBitCS \cite{PV2013} & xval & 90.1\% & 193 / \bf 1 & \bf 1.3\%
        \end{tabular}
        \caption{A comparison of different classifiers on the ``swapped'' RCV1 binary dataset (where $M\gg N$), showing the test-set classification accuracy, the total and post-tuning runtimes, and the density of the weight vector.  Above, sp = sum-product; ms = max-sum; BG = Bernoulli-Gaussian; PR = Probit; HL = Hinge loss; L1 = $\ell_1$ regularization; LR = Logistic.}
        \label{tab:rcv1_performance_flipped}
\color{black}
\end{table}

\tabref{rcv1_performance_flipped} shows all \textb{$8$} classifiers achieving nearly identical test-set classification accuracy, \textb{with the exception of cross-validated OneBitCS, which gives noticeably poorer accuracy.
Interestingly, OneBitCS also gives by far the sparsest weight vectors, apparently at the cost of test-error rate.
A better tradeoff between test accuracy and weight vector density is given by the EM-tuned GAMP algorithms, which return weight vectors that are about half as dense as those returned by CDN, TRON, TFOCS, and cross-validated GAMP, but that sacrifice only a fraction-of-a-percent in test accuracy.}

 
\tabref{rcv1_performance_flipped} also shows a wide range of runtimes.
\textb{OneBitCS gives by far the fastest post-tuning runtime, for the reasons described earlier.  
Among the total runtimes, however, the two} fastest are EM-GAMP based, with the best (at \textb{$105$} seconds) beating the fastest \textb{high-accuracy} non-GAMP algorithm \textb{(i.e., TFOCS} at $1086$ seconds) by more than a factor of \textb{$10$}.
That said, some caution must be used when comparing runtimes.
For example, while all algorithms were given a ``stopping tolerance'' of $10^{-3}$, the algorithms apply this tolerance in different ways.
Also, CDN and TRON are implemented in C++, while GAMP is implemented in object-oriented MATLAB (and therefore is far from optimized).

\textb{To understand how performance is impacted in a data-starved regime (i.e., $N > M$), we tested each algorithm on the same RCV1 dataset, but \emph{without} swapping the train/test datasets as was done in \cite{LWK2007,YCHL2010} and our \tabref{rcv1_performance_flipped}.  
The results are shown in \tabref{rcv1_performance_nonflipped}.
Similar to our other RCV experiment, we see all classifiers yielding very similar test error rates, with the exception of OneBitCS, which does significantly worse.
Again, however, OneBitCS generates an extremely sparse weight vector at the expense of test error rate, whereas some the EM-tuned BG-HL and L1-LR GAMP algorithms offer (milder) density reduction without a significant cost in test accuracy.
Finally, the two fastest total runtimes are earned by the spGAMP algorithms, and the fastest (BG-PR at 4 seconds) is about 3 times as quick as the fastest high-accuracy non-GAMP algorithm (i.e., CDN at 11 seconds).}
\begin{table}[t]
\color{blue}
        \centering
        \begin{tabular}{@{}c@{~}c|c@{~}c@{~}c@{}}
                Classifier & Tuning & Accuracy & Runtime (s) & Density  \\
                \hline
                spGAMP:\,BG-PR & EM & 95.5\% & {\bf 4} / 4 & 7.4\% \\
                spGAMP:\,BG-HL & EM & 95.1\% & 6 / 6 & 3.1\% \\
                msGAMP:\,L1-LR & EM & 95.6\% & 16 / 3 & 1.8\%   \\
                msGAMP:\,L1-LR & xval & 95.5\% & 134 / 16 & 4.6\%      \\
                
                \hline

                CDN \cite{YCHL2010} & xval & 95.5\% & 11 / 2 & 5.0\% \\
                TRON \cite{LM1999} & xval & \bf 96.0\% & 19 / 3 & 12.4\% \\
                
                TFOCS \cite{BCG2011} & xval & 95.7\% & 17 / 2 & 4.3\% \\
                OneBitCS \cite{PV2013} & xval & 89.7\% & 8 / \bf 0.1 & \bf 0.8\%
        \end{tabular}
        \caption{\textb{A comparison of different classifiers on the ``non-swapped'' RCV1 binary dataset (with $N > M$), showing the test-set classification accuracy, the total and post-tuning runtimes, and the density of the weight vector.  Above, sp = sum-product; ms = max-sum; BG = Bernoulli-Gaussian; PR = Probit; HL = Hinge loss; L1 = $\ell_1$ regularization; LR = Logistic.}}
        \label{tab:rcv1_performance_nonflipped}
\color{black}
\end{table}

Finally, we note that, although GAMP was derived under the assumption that the elements of $\vec{X}$ are realizations of a an i.i.d sub-Gaussian distribution, it worked well even with the $\vec{X}$ of this experiment, which was far from i.i.d sub-Gaussian.
We attribute the robust performance of GAMP to the ``damping'' mechanism included in the GAMPmatlab implementation, \textb{which was first described in \cite{SR2012} and rigorously analyzed in \cite{RSF2014}.  
Essentially, damping slows down the updates with the goal of preventing divergence.}

\subsection{Robust Classification}
\label{sec:numerical_study:robust}
In \secref{gamp_classification:robust}, we proposed an approach by which GAMP can be made robust to labels that are corrupted or otherwise highly atypical under a given activation model $p^*_{\Y|\Z}$.  We now evaluate the performance of this robustification method.
To do so, we first generated examples\footnote{%
Data was generated according to a class-conditional Gaussian distribution with $N$ discriminatory features.  Specifically, given the label $y\in\{-1,1\}$ a feature vector $\vec{x}$ was generated as follows: entries of $\vec{x}$ were drawn i.i.d $\mathcal{N}(y\mu, M^{-1})$ for some $\mu>0$.  Under this model, with balanced classes, the Bayes error rate can be shown to be $\eps_{\text{B}} = \Phi(-\sqrt{NM}\mu)$.  The parameter $\mu$ can then be chosen to achieve a desired $\eps_{\text{B}}$.
} $(y_m,\vec{x}_m)$ with balanced classes such that the Bayes-optimal classification boundary is a hyper-plane with a desired Bayes error rate of $\eps_{\text{B}}$. 
Then, we flipped a fraction $\gamma$ of the training labels (but not the test labels), trained several different varieties of GAMP classifiers, and measured their classification accuracy on the test data.

The first classifier we considered paired a genie-aided ``standard logistic'' activation function, \eqref{logistic_model}, with an i.i.d. zero-mean, unit-variance Gaussian weight vector prior.  Note that under a class-conditional Gaussian generative distribution with balanced classes, the corresponding activation function is logistic with scale parameter $\alpha = 2M\mu$ \cite{J1995}.  Therefore, the genie-aided logistic classifier was provided the true value of $\mu$, which was used to specify the logistic scale $\alpha$.  The second classifier we considered paired a genie-aided robust logistic activation function, which possessed perfect knowledge of both $\mu$ and the mislabeling probability $\gamma$, with the aforementioned Gaussian weight vector prior.  To understand how performance is impacted by the parameter tuning scheme of \secref{parameter_tuning}, we also trained EM variants of the preceding classifiers.  The EM-enabled standard logistic classifier was provided a fixed logistic scale of $\alpha = 100$, and was allowed to tune the variance of the weight vector prior.  The EM-enabled robust logistic classifier was similarly configured, and in addition was given an initial mislabeling probability of $\gamma^0 = 0.01$, which was updated according to \eqref{gamma_em_update2}.

In \figref{robust_vs_standard}, we plot the test error rate for each of the four GAMP classifiers as a function of the mislabeling probability $\gamma$.  For this experiment, $\mu$ was set so as to yield a Bayes error rate of $\eps_{\text{B}} = 0.05$.  $M = 8192$ training examples of $N = 512$ training features were generated independently, with the test set error rate evaluated based on $1024$ unseen (and uncorrupted) examples.  Examining the figure, we can see that EM parameter tuning is beneficial for both the standard and robust logistic classifiers, although the benefit is more pronounced for the standard classifier.  Remarkably, both the genie-aided and EM-tuned robust logistic classifiers are able to cope with an extreme amount of mislabeling while still achieving the Bayes error rate, thanks in part to the abundance of training data.

\begin{figure}
	\begin{center}
		\ifthenelse{\boolean{ONE_COLUMN}}
		{\scalebox{0.35}{\includegraphics*[-0.5in,1.55in][8.75in,9.5in]{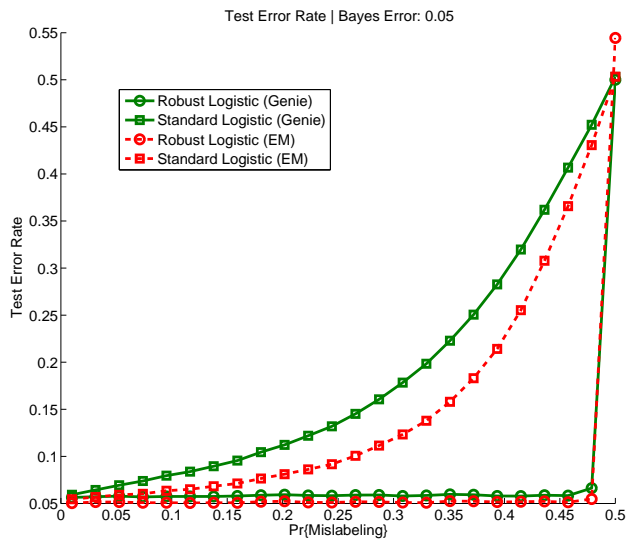}}}
		{\scalebox{0.35}{\includegraphics*[-0.5in,1.55in][8.75in,9.5in]{Figures/08Jul2013_235454266_fixed_bayes_err.eps}}}
		\caption{Test error rate of genie-aided (solid curves) and EM-tuned (dashed curves) instances of standard logistic (\Square) and robust logistic ($\circ$) classifiers, as a function of mislabeling probability $\gamma$, with $M = 8192$, $N = 512$, and Bayes error rate $\eps_{\text{B}} = 0.05$.}
		\label{fig:robust_vs_standard}
	\end{center}
\end{figure}

\subsection{Multi-Voxel Pattern Analysis}
\label{sec:numerical_study:mvpa}

Multi-voxel pattern analysis (MVPA) has become an important tool for analyzing functional MRI (fMRI) data \cite{HGFISP2001,NPDH2006,PMB2009}.  
Cognitive neuro-scientists, who study how the human brain functions at a physical level, employ MVPA not only to infer a subject's cognitive state but to gather information about how the brain itself distinguishes between cognitive states.
In particular, by identifying \emph{which} brain regions are most important in discriminating between cognitive states, they hope to learn the underlying processes by which the brain operates.
In this sense, the goal of MVPA is often feature selection, not classification.



To investigate the performance of GAMP for MVPA, we conducted an experiment using the well-known Haxby dataset \cite{HGFISP2001}. 
The Haxby dataset consists of fMRI data collected from $6$ subjects with $12$ ``runs'' per subject. 
In each run, the subject passively viewed blocks of $9$ greyscale images from each of $8$ object categories (i.e., faces, houses, cats, bottles, scissors, shoes, chairs, and nonsense patterns), 
during which full-brain fMRI data was recorded
over $N=31\,398$ voxels.

In our experiment, we designed classifiers that predict binary object category (e.g., cat vs.~scissors) from $M$ examples of $N$-voxel fMRI data collected from a single subject.
For comparison, we tried \textb{four} algorithms:
i) $\ell_1$-penalized logistic regression (L1-LR) as implemented using cross-validation-tuned TFOCS \cite{BCG2011},
ii) L1-LR as implemented using EM-tuned max-sum GAMP, 
iii) sum-product GAMP under a Bernoulli-Laplace prior and logistic activation function (BL-LR)\textb{, and iv) a cross-validation-tuned OneBitCS \cite{PV2013} classifier}.

Algorithm performance (i.e., error-rate, sparsity, and consistency) was assessed using $12$-fold leave-one-out cross-validation. 
In other words, for each algorithm, $12$ separate classifiers were trained, each for a different combination of $1$ testing fold (used to evaluate error-rate) and $11$ training folds. 
The reported performance then represents an average over the $12$ classifiers.
Each fold comprised one of the runs described above, and thus contained $18$ examples (i.e., $9$ images from each of the $2$ object categories constituting the pair), yielding a total of $M=11\times 18=198$ training examples.
Since $N=31\,398$, the underlying problem is firmly in the $N\gg M$ regime.

To tune each TFOCS classifier (i.e., select its $\ell_1$ regularization weight $\lambda$), we used a second level of leave-one-out cross-validation. 
For this, we first chose a fixed $G\!=\!10$-element grid of logarithmically spaced $\lambda$ hypotheses.
Then, for each hypothesis, we designed $11$ TFOCS classifiers, each of which used $10$ of the $11$ available folds for training and the remaining fold for error-rate evaluation.
Finally, we chose the $\lambda$ hypothesis that minimized the error-rate averaged over these $11$ TFOCS classifiers.  
\textb{A similar two-level cross-validation strategy was applied for selection of the sparsity rate in OneBitCS, using a logarithmically spaced $50$-point grid over sparsity rates between $0.1$\% and $15$\%.}
For EM-tuned GAMP, there was no need to perform the second level of cross-validation:  
we simply applied the EM tuning strategy described in \secref{parameter_tuning} to the $11$-fold training data.

\begin{table*}[t]
\color{blue}
        \centering
        \newcommand{\spc}{\hspace{1mm}}
        \newcommand{\sz}{\ssmall}
        \begin{tabular}{c|c@{\spc}c@{\spc}c@{\spc}c|c@{\spc}c@{\spc}c@{\spc}c|c@{\spc}c@{\spc}c@{\spc}c|c@{\spc}c@{\spc}c@{\spc}c}
                & \multicolumn{4}{c|}{Error Rate (\%)} &\multicolumn{4}{c|}{Sparsity (\%)} & \multicolumn{4}{c|}{Consistency (\%)} & \multicolumn{4}{c}{Runtime (s)}    \\
                Comparison & 
                        \sz TFOCS & \sz L1-LR & \sz BG-LR & \sz 1-Bit & 
                        \sz TFOCS & \sz L1-LR & \sz BG-LR & \sz 1-Bit & 
                        \sz TFOCS & \sz L1-LR & \sz BG-LR & \sz 1-Bit & 
                        \sz TFOCS & \sz L1-LR & \sz BG-LR & \sz 1-Bit \\
                \hline
                Cat vs. Scissors & 
                        9.7 & 11.1 & 9.3 & \textbf{5.1} & 
                        0.1 & 0.07 & \textbf{0.01} & 0.12 & 
                        38 & 43 & \textbf{60} & 57 & 
                        1318 & 137 & \bf 21 & 202 \\
                Cat vs. Shoe & 
                        \textbf{6.1} & \textbf{6.1} & 11.6 & 6.5 & 
                        0.14 & 0.07 & \textbf{0.01} & 0.12 & 
                        34 & 47 & \textbf{60} & 59 & 
                        1347 & 191 & \textbf{24} & 205 \\
                Cat vs. House & 
                        0.4 & \textbf{0.0} & 1.4 & 3.7 & 
                        0.04 & 0.02 & \textbf{0.01} & 0.12 & 
                        53 & \textbf{87} & 84 & 75 & 
                        1364 & 144 & \textbf{18} & 202 \\
                Bottle vs. Shoe & 
                        29.6 & 30.5 & 23.6 & \textbf{20.4} & 
                        0.2 & 0.1 & \textbf{0.01} & 0.12 & 
                        23 & 31 & 36 & \textbf{53} & 
                        1417 & 166 & \bf 22 & 205   \\
                Bottle vs. Chair & 
                        \textbf{13.9} & \textbf{13.9} & 15.7 & 26.9 & 
                        0.1 & 0.07 & \textbf{0.01} & 0.12 & 
                        30 & 45 & \textbf{61} & 37 & 
                        1355 & 150 & \bf 21 & 203 \\
                Face vs. Chair & 
                        \textbf{0.9} & \textbf{0.9} & 6.9 & 2.8 & 
                        0.09 & 0.05 & \textbf{0.01} & 0.12 & 
                        43 & 67 & 68 & \textbf{76} & 
                        1362 & 125 & \bf 24 & 205 \\
                \hline
                Average &
                        \bf 10.1 & 10.4 & 11.4 & 10.9 & 
                        0.11 & 0.06 & \bf 0.01 & 0.12 & 
                          37 &   53 & \bf 62   & 60 & 
                        1358 & 152  & \bf 22   & 204  
        \end{tabular}
        \caption{\textb{Performance of cross-validation tuned L1-LR TFOCS (``TFOCS''), EM-tuned L1-LR max-sum GAMP (``L1-LR''), EM-tuned BG-LR sum-product GAMP (``BG-LR''), and cross-validation tuned OneBitCS (``1-Bit'') classifiers on various Haxby pairwise comparisons.}}
        \label{tab:haxby_perf_table}
\color{black}
\end{table*}

\tabref{haxby_perf_table} reports the results of the above-described experiment for six pairwise comparisons.
\textb{For all but BG-LR GAMP, sparsity refers to the average percentage of non-zero
elements in the learned weight vectors. 
But, since BG-LR GAMP's weights are non-zero with probability one, we instead define BG-LR's sparsity as the number of weights with posterior probability $p(w_n\!\neq\! 0|\vec{y})>1/2$, as we did with the other sum-product-GAMP classifers in earlier experiments.}
Consistency refers to the average Jaccard index between weight-vector supports, i.e.,
\begin{equation}
        \text{consistency} 
        \defn \frac{1}{12}\sum_{i=1}^{12} \frac{1}{11}\sum_{j\neq i}\frac{|\mathcal{S}_i \cap \mathcal{S}_j|}{|\mathcal{S}_i \cup \mathcal{S}_j|}
\end{equation}
where $\mathcal{S}_i$ denotes the support of the weight vector learned when holding out the $i^{th}$ fold.
Runtime refers to the total time used to complete the $12$-fold cross-validation procedure.

\textb{Ideally, we would like an algorithm that quickly computes weight vectors with low estimated error rate, high consistency, and relatively low density. 
It should be emphasized that minimizing estimated error rate alone is not of sole importance, especially for this dataset, where the total number of samples is so few that the error rate estimates are understood to be very noisy.
Moreover, since the goal of MVPA is to identify which voxels of the brain are most important in discriminating between cognitive states, consistency among folds \emph{is} very important.}

\textb{Unfortunately, \tabref{haxby_perf_table} reveals no clear winner among the algorithms under test.
Starting with the estimated error rates, all four algorithms yielded similar comparison-averaged rates, spanning the range from $10.1$\% (for TFOCS) to $11.4$\% (for BG-LR GAMP).
Interestingly, the algorithm ranking under the consistency metric was exactly the opposite of that for the error-rate metric: BG-LR GAMP yielded the highest consistency (of $62$\%) and TFOCS the lowest consistency (of $37$\%).
In terms of sparsity, BG-LR GAMP appears to be the winner, but perhaps a direct comparison to the other algorithms should be avoided due to the differences in the definition of sparsity.
For runtime, however, the clear winner is EM-tuned BG-LR GAMP, which runs an order-of-magnitude faster than cross-validated OneBitCS and nearly two orders-of-magnitude faster than cross-validated TFOCS.}

\textb{A direct comparison between cross-validated TFOCS and EM-tuned L1-LR GAMP is illuminating, since these two algorithms share the L1-LR objective and thus differ mainly in tuning strategy.\footnote{\textb{It is known that, if max-sum GAMP converges, then it converges to a critical point of the optimization objective \cite{RSRFC2013}, which in the (convex) L1-LR case is unique.}} 
For this Haxby data, \tabref{haxby_perf_table} shows that L1-LR GAMP's classifiers are uniformly more sparse (and nearly twice as sparse on average) as those generated by TFOCS, while suffering only a small degradation in error-rate.
Meanwhile, L1-LR GAMP's classifiers are uniformly more consistent, and its runtimes are about $9\times$ faster on average.}

\section{Conclusion}
\label{sec:conclusion}
In this work, we presented the first comprehensive study of the \emph{generalized approximate message passing} (GAMP) algorithm \cite{R2011} in the context of linear binary classification \textb{and feature selection}.  We established that a number of popular discriminative models, including logistic and probit regression, as well as support vector machines (via hinge loss), can be implemented in an efficient manner using the GAMP algorithmic framework, and that GAMP's state evolution formalism can be used in certain instances to predict the misclassification rate of these models.  In addition, we demonstrated that a number of sparsity-promoting weight vector priors can be paired with these activation functions to encourage feature selection.  Importantly, GAMP's message passing framework enables us to learn the hyper-parameters that govern our probabilistic models adaptively from the data using expectation-maximization (EM), a trait which can be advantageous \textb{in terms of runtime}.  The flexibility imparted by the GAMP framework allowed us to 
consider several modifications to the basic discriminative models, such as robust classification, which can be effectively implemented using existing non-robust modules.  

In a numerical study, we confirmed the efficacy of our approach on both synthetic and real-world classification problems.  
For example, we found that the proposed EM parameter tuning can be both computationally efficient and accurate in the applications of text classification and multi-voxel pattern analysis.  
We also observed on synthetic data that \textb{GAMP can attain nearly optimal error rates in the $N\gg M$ regime when $N$ is sufficiently large and the number of discriminatory features, $K$ is sufficiently small.  Furthermore, we observed that} the robust classification extension can substantially outperform a non-robust counterpart.  

\appendices

\section{Sum-Product GAMP Hinge-Loss Computations}
\label{app:sum_product_hinge}
\allowdisplaybreaks
In this appendix, we describe the steps needed to compute the sum-product GAMP nonlinear steps for the hinge-loss activation function, \eqref{hinge_prior}.  For convenience, we define the associated \emph{un-normalized} likelihood function
\begin{eqnarray}
        \tilde{p}_{\Y|\Z}(y|z) &\triangleq& \exp\big(-\max(0, 1 - y z)\big),    \quad y \in \{-1,1\}.
        \label{eq:hinge_prior2}
\end{eqnarray}
Note from \eqref{z_posterior} that the sum-product $(\hat{z},\tau_z)$ can be interpreted as the posterior mean and variance of a random variable, $\Z$, with prior $\mathcal{N}(\hat{p}, \tau_p)$ and 
likelihood proportional to $\tilde{p}_{\Y|\Z}(y|z)$.

To compute the statistics $\hat{z} \equiv \text{E}[\Z|\Y\!=\!y]$ and $\tau_z \equiv \text{var}\{\Z|\Y\!=\!y\}$, we first write the posterior pdf as
\begin{equation}
        p_{\Z|\Y}(z|y) = C_y^{-1} \tilde{p}_{\Y|\Z}(y|z) p_{\Z}(z),
        \label{eq:hinge_z_posterior}
\end{equation}
where $C_y$ is an appropriate normalization constant.  Defining 
\begin{align}
\alpha_y &\triangleq ((1 - \tau_p) - y\hat{p})/\sqrt{\tau_p} \\ 
\beta_y &\triangleq (y\hat{p} - 1)/\sqrt{\tau_p} \\ 
\delta_y &\triangleq y\hat{p} - 1 + \tau_p/2, 
\end{align}
it can be shown \cite{Z2014} that 
\begin{align}
        C_1     &= \int_{-\infty}^{1} \exp(z - 1) \mathcal{N}(z; \hat{p}, \tau_p) + \int_{1}^{\infty} \mathcal{N}(z; \hat{p}, \tau_p)   \label{eq:c1_s1} \\
                &= \exp(\delta_1) \Phi(\alpha_1) + \Phi(\beta_1)
                \label{eq:const1_integral}
\end{align}
The posterior mean for $y=1$ is therefore given by
\begin{eqnarray}
        \lefteqn{\text{E}[\Z|\Y\!=\!1]
                = \frac{1}{C_1} \int_z z \, \tilde{p}_{\Y|\Z}(y\!=\!1|z) p_{\Z}(z)  }\\
                &=& \frac{1}{C_1} \left[ e^{\delta_1} \int_{-\infty}^{1} \!\! z \mathcal{N}(z; \hat{p} + \tau_p, \tau_p) 
                + \int_{1}^{\infty} \!\! z \mathcal{N}(z; \hat{p}, \tau_p) \right]      \\
                &=& \tfrac{e^{\delta_1} \Phi(\alpha_1)}{C_1} \int_{-\infty}^{1}  \!\!z \tfrac{\mathcal{N}(z; \hat{p} + \tau_p, \tau_p)}{\Phi(\alpha_1)} 
                + \tfrac{\Phi(\beta_1)}{C_1} \int_{1}^{\infty}  \!\!z \tfrac{\mathcal{N}(z; \hat{p}, \tau_p)}{\Phi(\beta_1)}, \quad
        \label{eq:mean1_s1}
\end{eqnarray}
where each integral in \eqref{mean1_s1} represents the first moment of a truncated normal random variable.  
Similar expressions can be derived for $\text{E}[\Z|\Y\!=\!-1]$.
Then, defining the quantities
\begin{align}
        \gamma_y &\triangleq e^{-\delta_y}\Phi(\beta_y)/\Phi(\alpha_y)  \\
        \ubar{\mu}_y &\triangleq \hat{p} + y \big(\tau_p - \sqrt{\tau_p} \phi(\alpha_y)/\Phi(\alpha_y)\big)     \\
        \bar{\mu}_y &\triangleq \hat{p} + y \sqrt{\tau_p} \phi(\beta_y)/\Phi(\beta_y),
\end{align}
it can be shown \cite{BS1999} that, for $y\in\{-1,1\}$, 
\begin{equation}
        \hat{z}(y) = \text{E}[\Z|\Y=y] = (1 + \gamma_y)^{-1} \ubar{\mu}_y + (1 + \gamma_y^{-1})^{-1} \bar{\mu}_y.
        \label{eq:hinge_mean}
\end{equation}

To compute $\tau_z \equiv \text{var}\{\Z|\Y=y\}$, it suffices to derive an expression for $\text{E}[\Z^2|\Y=y]$.  Following the same line of reasoning that produced \eqref{mean1_s1}, we find
\begin{eqnarray}
        \lefteqn{ \text{E}[\Z^2|\Y\!=\!1] }\label{eq:secondmoment1_s1}\\
        &=& \tfrac{e^{\delta_1} \Phi(\alpha_1)}{C_1} \!\int_{-\infty}^{1} \!\!\! z^2 \tfrac{\mathcal{N}(z; \hat{p} + \tau_p, \tau_p)}{\Phi(\alpha_1)} 
        + \tfrac{\Phi(\beta_1)}{C_1} \!\int_{1}^{\infty} \!\!\! z^2 \tfrac{\mathcal{N}(z; \hat{p}, \tau_p)}{\Phi(\beta_1)}, \nonumber
\end{eqnarray}
where each integral in \eqref{secondmoment1_s1} is the second moment of a truncated normal random variable. 
A similar expression can be derived for $\text{E}[\Z^2|\Y\!=\!-1]$.
Defining
\begin{eqnarray}
        \ubar{v}_y &\triangleq& \tau_p \left[1 - \frac{\phi(\alpha_y)}{\Phi(\alpha_y)} \left(\frac{\phi(\alpha_y)}{\Phi(\alpha_y)} + \alpha_y \right) \right]    \\
        \bar{v}_y &\triangleq& \tau_p \left[1 - \frac{\phi(\beta_y)}{\Phi(\beta_y)} \left(\frac{\phi(\beta_y)}{\Phi(\beta_y)} + \beta_y \right) \right],
\end{eqnarray}
it can be shown \cite{BS1999} that
\begin{equation}
        \text{E}[\Z^2 | \Y=y] = (1 + \gamma_y)^{-1} (\ubar{v}_y + \ubar{\mu}_y^2) + (1 + \gamma_y^{-1})^{-1} (\bar{v}_y + \bar{\mu}_y^2),
\end{equation}
allowing us to compute $\tau_z(y) = \text{E}[\Z^2 | \Y=y] - \hat{z}^2(y)$.

\bibliographystyle{ieeetr}
\bibliography{gamp_classification}

\end{document}